 \documentclass[10pt,twocolumn]{revtex4}

 \usepackage{amssymb}
\usepackage{graphicx}
\usepackage{caption}
\usepackage{amsmath}
\usepackage{bbold}
\usepackage{bm}

\newcommand{\HH}{\mathcal{H}}

\renewcommand{\b}[1]{\mathbf{ #1}}									

\newcommand{\rom}[1]{\uppercase\expandafter{\romannumeral #1\relax}} 

\newcommand{\h}[1]{\hat{#1}}										

\renewcommand{\Re}{\operatorname{Re}}

\begin{document}
       \title{Quantum simulators based on the global collective light-matter interaction}
        \author{Santiago F.  Caballero-Benitez, Gabriel Mazzucchi and Igor B. Mekhov}
      \affiliation{ 
 University of Oxford, Department of Physics, Clarendon Laboratory, Parks Road, Oxford OX1 3PU, UK}

\begin{abstract}
We show that coupling ultracold atoms in optical lattices to quantized modes of an optical cavity leads to quantum phases of matter, which at the same time posses properties of systems with both short- and long-range interactions. This opens perspectives for novel quantum simulators of finite-range interacting systems, even though the light-induced interaction is global (i.e. infinitely long range). This is achieved by spatial structuring of the global light-matter coupling at a microscopic scale. Such simulators can directly benefit from the collective enhancement of the global light-matter interaction and constitute an alternative to standard approaches using Rydberg atoms or polar molecules. The system in the steady state of light induces effective many-body interactions that change the landscape of the phase diagram of the typical Bose-Hubbard model. Therefore, the system can support non-trivial superfluid states, bosonic dimer, trimers, etc. states and supersolid phases depending on the choice of the wavelength and pattern of the light with respect to the classical optical lattice potential. We find that by carefully choosing the system parameters one can investigate diverse strongly correlated physics with the same setup, i.e., modifying the geometry of light beams. In particular, we present the interplay between the density and bond (or matter-wave coherence) interactions. We show how to tune the effective interaction length in such a hybrid system with both short-range and global interactions. 
\end{abstract}

       \maketitle

\section{Introduction}

Ultracold gases loaded in optical lattice is the ideal tool for studying the quantum degenerate regime of matter. Controlling the coupling between the atoms and the light beams creating the optical lattice allows to realize simple models~\cite{Lewenstein} that were first formulated in different fields of physics from condensed matter to particle physics and  biological systems. These models would be useful for quantum simulation purposes~\cite{Bloch} and quantum information processing (QIP) applications. Specifically, one can realize effective Hamiltonians which contain short-range physical processes such as tunneling between neighbor lattice sites and on-site interactions. The implementation of long-range interactions that extends over many lattice sites is an extremely challenging task since it requires the use of more complex systems such as polar molecules~\cite{lahaye2009,Ferlaino} or Rydberg atoms~\cite{pohl2009,PohlLukin,Fleischhauer}. Moreover, spatial structure of the interaction itself is fixed by the physical system used (e.g. dipole-dipole interaction for molecules and Van der Waals interaction for Rydberg atoms) and cannot be changed. 

 In contrast to these examples, we show that by loading an optical lattice inside a cavity allows to engineer synthetic many-body interactions with an arbitrary spatial profile. These interactions are mediated by the light field and do not depend on fundamental processes, making them extremely tunable and suitable for realizing quantum simulations of many-body long-range Hamiltonians. In contrast to other proposals based on light-mediated interactions~\cite{Porras2006,Strack2011,IonsFR2012, Lesanovsky2013,PhotCrys2015}, we suggest a novel approach, where the shortening of the a priori infinitely long-range (global) light-induced interaction does not degrade the collective light-matter interaction, but rather benefits from it. In particular, in contrast to other proposals, where shortening of the interaction length requires increasing number of light modes, in our case, rather short-range interactions can be simulated with small number of light modes. Moreover, even a single mode cavity is enough to simulate some finite-range interactions. As a result, the quantum phase of matter posses properties of systems with both short-range and global collective interactions. The effective Hamiltonians can be an acceptable representation of an otherwise experimentally hard to achieve quantum degenerate system with finite range interaction.

In this article we present how different arrangements involving multiple probes and/or multiple light modes for detecting the scattered light, lead to these synthetic interactions. The setups we propose are an extension of recent experimental breakthroughs where optical lattices in a single mode cavity have been achieved~\cite{Hemmerich2015,Esslinger2015}. Under these conditions, light and matter are both in the full quantum regime, thus we have a quantum optical lattice. This behaviour follows from the dynamical properties of the light~\cite{RitschRMP} and the structural Dicke phase transition that occurs and forms a state with supersolid features~\cite{EsslingerNat2010}.  Currently, the study of the full quantum regime of the system has been limited to few atoms~\cite{EPJD08,VukicsNJP2007,RitschLight,KramerRitschPRA2014, RitschArxiv2015}. As the light matter coupling is strongly enhanced in a high finesse optical cavity in a preferred wavelength, the atoms reemit light with the backaction comparable with that of the lasers used in the trapping process. As a consequence, an effective long-range (nonlocal) interaction emerges driven by the cavity field. It is now experimentally possible to access the regime where light-matter coupling is strong enough and the cavity parameters allow  to study the formation of quantum many-body phases with cavity decay rates of MHz~\cite{PNASEsslinger2013, Esslinger2015} and kHz~\cite{PNASHemmerich2015, Hemmerich2015}. The light inside the cavity can be used to control the formation of many-body phases of matter even in a single cavity mode~\cite{EPJD08,MekhovNP2007,MekhovRev,PRL2015, NJPhys2015}. This leads to several effects yet to be observed due to the dynamical properties of light~\cite{Larson,Larson2,Morigi,Hofstetter,Reza,Chin,Doner,Morigi2}. Moreover, it has been shown that multimode atomic density patterns can emerge, even their coherences can become structured and light-matter quantum correlations can control the formation of correlated phases. { Our goal is to explore the landscape of emergent quantum many-body phases the system is able to support by means of these light induced interactions. We will show a representative sample of some of the non-trivial quantum many body phases that can be achieved in experiments and their characteristics. } When light scatters maximally the system in its steady state has a dynamical optical lattice (DOL), while when the atoms scatter light minimally the system has a quantum (QOL)~\cite{PRL2015}. In a DOL the self-consistent nature of the matter-cavity system leads to structured self-organized states. In a QOL atomic quantum fluctuations are modified via the quantum fluctuations of the light field triggering the formation or stabilization of correlated phases. Recent experimental achievements~\cite{Hemmerich2015,Esslinger2015} are due to the dynamical nature of the OL via cavity backaction. Thus, a plethora of novel quantum phases due to the imprinting of structure by design in the effective light-induced interaction occurs~\cite{PRL2015, NJPhys2015,BO2016}. In addition quantum many-body phases can be measured via light-scattering measurements~\cite{NJPhys2015,MekhovNP2007, QNDSanpera,PRLMekhov09,LP09,Wojciech,Ueda}, or by matter wave scattering~\cite{HPM4}, and dynamical structure factors can be obtained via homodyne detection~\cite{structureFMW} and leaking of photons from the cavity~\cite{structureEsslinger}. Recently, density ordering has been achieved with classical atoms~\cite{LabeyrieNaturePhotonics2014}. Also, non interacting fermions in a cavity have been studied \cite{Fermions1, Fermions2, Fermions3} and even chiral states are possible~\cite{CFermions}.  Additionally, multimode cavities extend the possibilities for quantum phases even further~\cite{Strack2011,KramerRitschPRA2014,LevNaturePhys, Kollar2015,Muller2012}. Thus making our synthetic interactions feasible in the near future.  We will explore how by carefully tuning system parameters and the spatial structure of light, one can design with plenty of freedom the quantum many-body phases that emerge and one can even simulate finite range many-body interactions using the mode structure imprinted by the light. The quantum nature of the potential seen by the atoms will change the landscape of correlated quantum many-body phases beyond  classical optical lattice setups. 

Our work will foster the design of multicomponent states and their possible application towards quantum multimode systems in optomechanics~\cite{RMP2014Optomech}. Moreover, our approach can aid in the simulation by means of networks~\cite{QnetCirac,QnetBeige,QnetRempe}. Toward quantum state engineering, our models can be used for engineering non-trivial correlated many-body quantum states~\cite{GabrielAFM,WojciechB},
and quantum state preparation using state projection~\cite{PRA2009,LP10,LP11,LesanovskyDis, ZimmermanRing,Gabriel,Molmer,Vuletic,Gabriel2} via measurement back-action in addition to cavity back-action. Moreover feedback control~\cite{Zimmerman,Pedersen,Denis1,Denis2,Denis3,Denis4,Feedback} can also be explored in relation with the emergent phases we find. The our mode structures will appear in the realm of non-Hermitian dynamics via effective Hamiltonians~\cite{Diehl,TELee,Dhar,WojciechNHQZ} and polaritonic systems~\cite{Polaritons1,Polaritons2,Polaritons3}.

The setup we will describe might aid in the design of novel quantum materials, where the concepts we will show could be translated to real materials and composite devices in solid state systems.

 The article has the structure as follows. We introduce the model of ultracold atoms in high-Q cavity where the atoms are in the regime of quantum degeneracy and state the effective matter Hamiltonian that we will study. Then we show, how one can construct arbitrary interactions using the light induced structures that are formed. These effective interactions will be useful for the purpose of quantum simulation of finite range interactions among other possible applications. Then we show, how one can construct  and effective general representation of the mode Hamiltonians, and solve relevant cases for current and future experiments depending on the geometry of light. We will show the emergent quantum many-body phases that arise due to the effect of quantum light and some of their properties. We conclude our manuscript by discussing our results.

\section{The model}
\begin{figure}[htb!]
\centering
\includegraphics[width=0.47\textwidth]{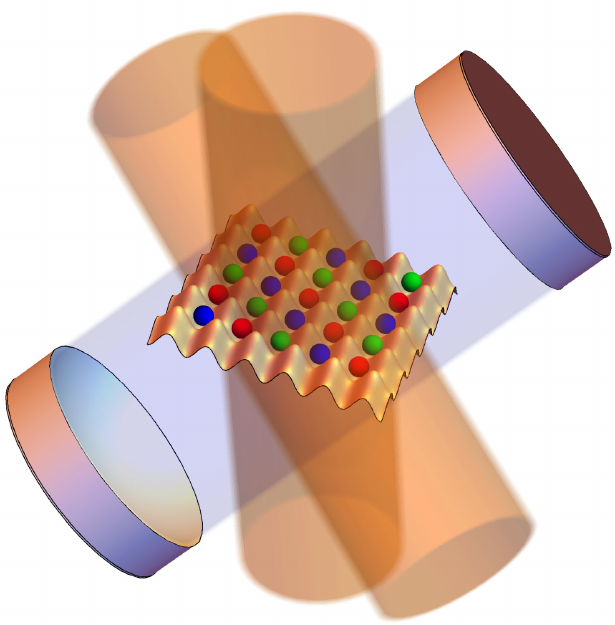}
\captionsetup{width=0.48\textwidth,justification=centerlast,font=small}
\caption{[color on-line] Setup of the system. The cold atoms in the optical lattice are inside a high reflective optical cavity, and several atomic modes are induced by light (spheres with different shades [colors] correspond to different atomic modes), while one or several coherent pumps are injected into the system.}\label{setup}
\end{figure}

The system consists of atoms trapped in an OL inside (a) single- or multi-mode cavity(ies) with the mode frequency(ies) $\omega_c$ and decay rate(s) $\kappa_c$ in off-resonant scattering~\cite{MekhovRev,Gabriel,Wojciech,PRL2015}.  {The off-resonant light scattering condition means that $\Gamma \ll |\Delta_{pa}|$, where $\Gamma$ is  the spontaneous emission rate of the atoms, where  $\Delta_{pa}=\omega_p-\omega_a$ is the detuning between the light mode(s)  frequency(ies) $\omega_p$ and the atomic resonance frequency $\omega_a$.}  The atomic system is probed with classical beam(s) and the scattered light is selected and enhanced by  the optical cavity(ies). The  light from the pump(s) has amplitude(s)  $\Omega_p$ (in units of the Rabi frequency). The cavity-pump detunning is $\Delta_{pc}=\omega_p-\omega_c$. The system is depicted in Fig.\ref{setup}.  The light is pumped from the side of the main axis of the high Q cavity(ies), at an angle not necessarily at $90^\circ$.   The cavity mode(s) couple(s) with the atoms via the effective coupling strength(s) $g_p= g_c \Omega_p/(2\Delta_{pa})$, with $g_c$ the light-matter coupling coefficient of the cavity(ies). The light-matter Hamiltonian describing the system is:
 $\HH=\HH^b+\HH^{a}+\HH^{ab}$, where $\HH^b$ is the regular Bose-Hubbard (BH) Hamiltonian~\cite{Fisher,Dieter},
\begin{equation}
\HH^b=-t_0\sum_{\langle i, j\rangle}(\hat b^\dagger_i\hat b^{\phantom{\dagger}}_j+\mathrm{H.c})-\mu\sum_i\hat n_i+\frac{U}{2}\sum_i\hat n_i(\hat n_i-1),
\end{equation}
with $t_0$ the nearest neighbor tunneling amplitude, $U$ the on-site interaction and $\mu$ the chemical potential.
The light is described by  $\HH^{a}=\sum_c\hbar\omega_c\hat a_c^\dagger \hat a^{\phantom{\dagger}}_c$ and the light-atom interaction is~\cite{MekhovRev}:
\begin{equation}
\HH^{ab}=\sum_{c,p} \left(g_p^*\hat a_c\hat F_{pc}^\dagger+g_p\hat a_c^\dagger\hat F_{pc}\right)
\label{LMp}
\end{equation}
with
$
\hat F_{pc}= \hat D_{pc}+\hat B_{pc}$. $\hat D_{pc}=\sum_{j}J^{pc}_{jj}\hat n_j$ is the density coupling of light to the atoms, $\hat B_{pc}=\sum_{\langle i,j\rangle}J_{ij}^{pc}( \hat b^\dagger_i\hat b^{\phantom{\dagger}}_j+h.c.)$ is due to the inter-site densities reflecting matter-field interference, or bonds~\cite{PRL2015,Wojciech}. The sums go over illuminated sites $N_s$, and nearest neighbor pairs $\langle i,j\rangle$. The operators $b_i^\dagger$ ($\hat b_i$) create (annihilate) bosonic atoms at site $i$, $\hat a^\dagger$  ($\hat a$)  photons in the cavity, while the number operator of atoms per site is given by $\hat n_i=\hat b_i^\dagger\hat b_i^{\phantom{\dagger}}$. $\HH^{ab}$  is the relevant contribution to the quantum potential seen by atoms on top of classical OL described by the BH model, where the on-site interaction $U$ and hopping amplitude $t_0$ are short-range local processes.
{  The effective parameters of the Bose-Hubbard Hamiltonian with the cavity field can be calculated from the Wannier functions and are given by
\begin{equation}
t_0=\int w(\mathbf{x}-\mathbf{x}_i)(\nabla^2-V_{OL}(\mathbf{x}))w(\mathbf{x}-\mathbf{x}_j)\mathrm{d}^n x,
\end{equation}
where $\langle i,j\rangle$ are nearest neighbor sites, and
\begin{equation} 
J^{pc}_{ij}=\int w(\mathbf{x}-\mathbf{x}_i)u^*_c(\mathbf{x})u_p(\mathbf{x})w(\mathbf{x}-\mathbf{x}_j)\mathrm{d}^n x,
\end{equation}
and ``$i$" and "$j$" can be the same site  for density coupling or be nearest neighbors for bond coupling, where $u_{c,p}(\mathbf{x})$ are the cavity(ies) and pump(s) mode functions and $w(\mathbf{x})$ are the Wannier functions. The classical optical lattice potential is given by $V_{OL}(\mathbf{x})$}. The light couples either to the density on each site or to the inter-site density, coupling the coherences.
The classical optical lattice defining the regular Bose-Hubbard Hamiltonian is weakly dependent of the cavity parameters~\cite{Larson2,Morigi}. The atoms are mainly trapped by the strong classical lattice, which is created inside a cavity(ies) by the external laser beams. This external potential is insensitive to the quantum state of atoms. The light scattered into the cavity constitutes a quantum perturbation of the strong classical potential. This perturbation strongly depends on the many-body atomic state. Furthermore, the classical optical lattice and  light in the cavity can be detuned from each other. 

Moreover, it is useful to exploit the spatial structure of light as a natural basis to define atomic modes, as the coupling coefficients $J^{pc}_{ij}$ can periodically repeat in space~\cite{PRL2015,NJPhys2015, Gabriel, Thomas,Wojciech}, and as we will show arbitrarily designed.  All atoms equally coupled to light belong to the same mode, while the ones coupled differently belong to different modes $\varphi$. Then we have for the atomic operators,
\begin{equation}
\hat F_{pc}=\sum_{\varphi}J^{pc}_{D,\varphi}\hat N_\varphi+\sum_{\varphi'}J^{pc}_{B,\varphi'}\hat S_{\varphi'}
\end{equation}
where the light induced ``density"  $\hat N_\varphi$ and  ``bond"  $\hat S_{\varphi}$ mode operators, such that:
\begin{equation}
\hat N_\varphi=\sum_{i\in\varphi}\hat n_i ,\; \textrm{and}\; \hat S_{\varphi}=\sum_{\langle i,j\rangle\in\varphi}(\hat b_i^\dagger\hat b_j^{\phantom{\dagger}}+\hat b_j^\dagger\hat b_i^{\phantom{\dagger}}),
\end{equation}
with $J^{pc}_{D,\varphi}$ corresponding to the posible values of $J^{pc}_{ii}$ and $J^{pc}_{B,\varphi'}$ corresponding to $J^{pc}_{ij}$ where the pair $\langle i,j\rangle$ are nearest neighbors. These encompass the different sets of values taken by the Wannier overlap integrals $J^{cp}_{ij}$~\cite{PRL2015, Gabriel, Thomas,Wojciech} for each mode of the cavity system, either a single mode cavity with one pump and one cavity or  a multi-mode cavity, and even multiple cavities and multiple pumps.

\section{Light mediated synthetic atom-atom interaction}

The Hamiltonian as it is has a light and matter sector that has a complicated structure. However, in the good cavity limit $|\Delta_{pc}|\gg\kappa_c$ while $|\Delta_{pc}|\gg |U|$ and $|\Delta_{pc}|\gg |t_0|$ we have that the light field can be adiabatically eliminated.  In general, the problem of reconstruction of the effective matter Hamiltonian is hard, and there are several alternatives for this reconstruction as shown in~\cite{EPJD08,Morigi,NJPhys2015}. Thus, the light field becomes enslaved by the matter and vice versa, we refer to this as cavity back-action. This leads to the effective Hamiltonian:
\begin{equation}
\HH_{\mathrm{eff}}=\HH^b+\sum_c\sum_{p,q}\left(\frac{g_{\mathrm{eff}}^{qc}}{2}\hat F_{pc}^\dagger\hat F_{qc}^{\phantom{\dagger}}+\frac{(g_{\mathrm{eff}}^{qc})^*}{2}\hat F_{pc}^{\phantom{\dagger}}\hat F_{qc}^\dagger\right)
\label{feff}
\end{equation}
with the effective coupling strengths,
\begin{equation}
g_{\mathrm{eff}}^{qc}=\frac{g^*_cg_q}{\Delta_{qc}+i\kappa_c}.
\end{equation}
 The sum over ``$c$" goes over the cavity modes (for a multi-mode cavity/several cavities) and ``$p$" and ``$q$" go over the number of pumps.
The new terms are the by-product of  the cavity back-action of light and matter: an effective structured interaction that depends on the form of coefficients given by the effective interaction couplings $g_{\mathrm{eff}}^{qc}$ and the projections of light onto the matter matter system given by the spatial distribution of the constants $J_{D,\varphi}^{pc}$, $J_{B,\varphi}^{pc}$ corresponding to the light induced modes $\varphi$. Carefully analysis shows~\cite{NJPhys2015}, that beyond the adiabatic limit, additional terms modify the effective Hamiltonian due to the non-commuting nature between light induced processes  with the original matter Hamiltonian and higher order photon processes. In what follows we will always consider the adiabatic limit.  The new terms beyond BH Hamiltonian give the effective long-range light-induced interaction between density and bond modes that depend on geometry of the cavity modes and light pumps injected into the system. We can rewrite the new terms as,
\begin{widetext}
\begin{eqnarray}
\sum_c\sum_{p,q}\left(\frac{g_{\mathrm{eff}}^{qc}}{2}\hat F_{pc}^\dagger\hat F_{qc}^{\phantom{\dagger}}+\frac{(g_{\mathrm{eff}}^{qc})^*}{2}\hat F_{pc}^{\phantom{\dagger}}\hat F_{qc}^\dagger\right)&=&
\sum_{\varphi,\varphi'}\sum_c\sum_{p,q}\Big[\tilde\gamma^{D,D}_{\varphi,\varphi'}(c,p,q)\hat N_\varphi^{\phantom{*}}
\hat N_{\varphi'}^{\phantom{*}}
+
\tilde\gamma^{B,B}_{\varphi,\varphi'}(c,p,q)\hat S_\varphi^{\phantom{*}}\hat S_{\varphi'}^{\phantom{*}}
\nonumber
\\
&&\phantom{\sum_{\varphi,\varphi'}\sum_c\sum_{p,q}}+
\tilde\gamma^{D,B}_{\varphi,\varphi'}(c,p,q)(\hat N_\varphi^{\phantom{*}}\hat S_{\varphi'}^{\phantom{*}}
+\hat S_{\varphi'}^{\phantom{*}} \hat N_{\varphi}^{\phantom{*}})\Big],
\label{mdecomp}
\end{eqnarray}
\end{widetext}
with,
$
\tilde\gamma^{\mu,\nu}_{\varphi,\varphi'}(c,p,q)=[g_{\mathrm{eff}}^{qc}(J_{\mu,{\varphi}}^{pc})^* J^{qc}_{\nu,{\varphi'}}+\mathrm{c.c.}]/2$, $\{\mu,\nu\}\in\{D,B\}
$. Cast in the above form, it is clear that the  terms are similar to  density-density interactions, exchange interactions, and the last two terms to a combination of the action of both multi mode density and exchange. The different values of the effective interaction couplings  $\tilde\gamma$ between light induced spatial mode operators $\hat N_\varphi$ and $\hat S_\varphi$ will determine the emergent phases in the system.
The structure constants together with the effective interaction couplings $g_{\mathrm{eff}}^{qc}$ give an unprecedented ability to design an arbitrary interaction profile between the atoms. The new terms from Eq. (\ref{mdecomp}) provide the system with a plethora of new possibilities.   In principle  one can design arbitrary interactions patterns by choosing the geometry of the light: modifying angles, pump amplitudes, detunings and cavity parameters. Certainly, the potential of this to enrich cold atomic systems in terms of the simulation and design of interactions is vast. In what follows, we will discuss some simple examples of how the above Hamiltonian can be used to generate synthetic interactions for quantum simulation, i.e. the couplings between the atoms is not a consequence of fundamental physical processes but are mediated by the light field and depend on the geometrical arrangement of the probe beam and the optical cavity. 

\subsection{One probe and one cavity}

\begin{figure}[htb!]
\centering
\includegraphics[width=0.49\textwidth]{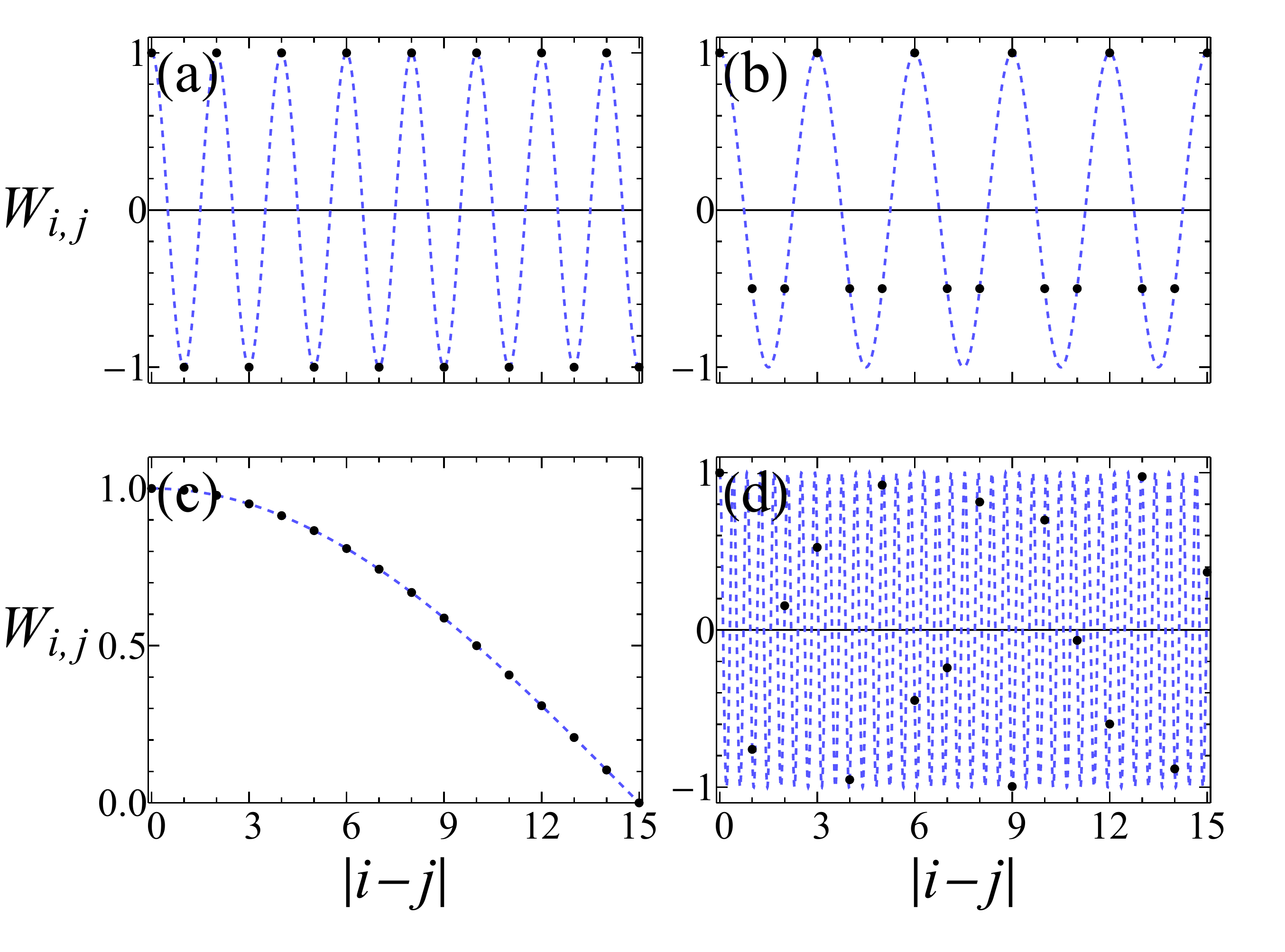}
\captionsetup{width=0.48\textwidth,justification=centerlast,font=small}
\caption{[color on-line] Different interactions that can be implemented using a single cavity mode and a single probe for $L=16$ lattice sites and using traveling waves as mode functions. The panels illustrate the value of $W_{i,j}$ as a function of $|i-j|$ (normalized). Panels (a) and (b): interaction strength in presence of $R=2$ ($\theta_{c_1}=0$, $\theta_{p_1}=\pi/2$, $\lambda=2a$) and $3$ ($\theta_{c_1}=0$, $\cos \theta_{p_1}=1/3$, $\lambda=2a$) spatial modes. The value of $W_{i,j}$ depends solely on the mode distance. Panel (c): increasing the number of atomic modes leads to long-range interactions with a well-defined spatial profile ($\theta_{c_1}=0$, $\cos \theta_{p_1}=7/8$, $\lambda=2a$). Panel (d): the periodicity of $W_{i, j}$ is incommensurate to the lattice spacing, resulting in a ``disordered'' interaction.}\label{fig:inter1}. The dashed lines are a guide to the eye between modes choosen.
\end{figure}

For a deep OL, the $\hat{B}_{cp}$ contribution to $\h{F}_{cp}$ can be neglected. 
We find that the effective atom-atom interaction  can be re-written as: 
\begin{align}\label{eq:Heff1}
\HH_1^D&=g_\mathrm{eff}\sum_{i,j}W_{ij} \h{n}_i \h{n}_j.
\end{align}
with $g_{\mathrm{eff}}=\Re[g_{\mathrm{eff}}^{11}]/2=\Delta_{11}|g_1|^2/(\Delta_{11}^2+\kappa_1^2)$, thus the effective interaction strength factors out. The specific dependence of the functions $J^{11}_{ii}$ defines the spatial profile of the interaction between the atoms via  the interaction matrix $W_{ij} = 2\Re \left[(J^{11}_{ii})^* J^{11}_{jj}\right]$. 
{
In order to illustrate this, we focus on a one-dimensional lattice and we consider traveling waves as mode functions for the light modes (i.~e. $u_c(\b{r})=e^{i \b{k}_c \cdot \b{r}}$ and $J^{11}_{jj}=e^{i (\b{k}_{c_1}-\b{k}_{p_1}) \cdot \b{r}_j}$) so that $W_{ij}$ becomes:
\begin{align}\label{eq:IntMat1}
W_{ij}=\cos \left[(\b{k}_{c_1}-\b{k}_{p_1})\cdot (\b{r}_i-\b{r}_j)\right].
\end{align}
The cavity induces a periodic interaction between the atoms and, depending on the projection of $\b{k}_{c_1}-\b{k}_{p_1}$ along the lattice direction $\hat e_{\b{r}_i}$, its spatial period can be tuned to be commensurate or incommensurate to the lattice spacing. Specifically, if $J^{11}_{jj}=e^{i 2 \pi j /R} $ and $R \in \mathbb{Z}^+$, atoms separated by $R$ lattice sites scatter light with the same phase and intensity and the optical lattice is partitioned in $R$ macroscopically occupied regions (spatial modes) composed by non-adjacent lattice sites \cite{Thomas}. The specific geometric configuration of the light beams is determined by the angles $\theta_{c_1,p_1}$ between the wave vectors $\b{k}_{c_1,p_1}$ and $e_{\b{r}_i}$: in order to have $R$ spatial modes, one has to set
\begin{align}\label{geometry1}
\cos \theta_{p_1}=\cos \theta_{c_1} - \frac{\lambda}{a} \frac{1}{R}
\end{align}
where $\lambda$ is the wavelength of the light modes.} If this relation is fulfilled, $W_{ij}$ has periodicity $R$ lattice sites and, defining the operator $\h{N}_j$ to be the population of the mode $j$, the interaction strength between two atoms belonging to the different modes $i$ and $j$ depends solely on their \emph{mode distance} ($(i-j) \mathrm{mod}R$) and not their actual separation. In this case, we find that the mode-mode interaction Hamiltonian is given by
\begin{align}\label{interaction1}
\HH^D_1=2 g_\mathrm{eff} \sum_{i,j=1}^R \cos \left(\frac{2 \pi}{R}(i-j)\right) \h{N}_i \h{N}_j.
\end{align}
The case $R=2$ has been recently realized \cite{Esslinger2015}, showing that new light-mediated interaction heavily affects the ground state of the Bose-Hubbard model inducing a new supersolid phase. Simply changing the angle between the cavity and the optical lattice, one can increase the number of spatial modes modes and therefore implement more complicated long-range interactions that cannot be obtained using molecules or Rydberg atoms [Fig.~\ref{fig:inter1}].

If the period of $W_{ij}$ is incommensurate to the lattice spacing, the spatial modes are not well-defined since each lattice sites scatters light with a different phase and amplitude. Therefore, the value of interaction matrix is no longer periodic and resembles disorder~\cite{Morigi2}. This behavior is analogous to the potential generated by the superposition of two incommensurate optical lattices~\cite{lewenstein2010} which has been used for generating controllable disordered potential, allowing the observation of Anderson localization and new quantum phases (Bose glass) in ultracold gases~\cite{roati2008,derrico2014}. Here, we move a step further and we create synthetic random interactions between the atoms, generalizing the Anderson model where disorder affects only the local potential and/or the tunneling amplitude~\cite{SantosUDis,UDis2015}.

\subsection{Multiple probes and one cavity} 
The previous simple case where only one cavity and one probe are present allows to realize interactions that resemble a cosine profile. We know turn to a different monitoring scheme where the atoms are probed with $R$ different classical pumps and scatter light to a single optical cavity.

In addition to the previous paragraphs, where the light mode of the cavity has contributions from all the sites of the optical lattice, all the probe beams concur to the value of the light field inside the cavity and the effective atom-atom interaction is 
\begin{align}\label{eq:Heff2}
\HH^D_1&=\sum_{p,q=1}^{R}\left(\frac{g^{q1}_\mathrm{eff}}{2}\h{D}_{p1}^\dagger \h{D}_{q1} + \mathrm{H.c.}\right)
\end{align}
\begin{figure*}[htb!]
\centering
\includegraphics[width=1.03\textwidth]{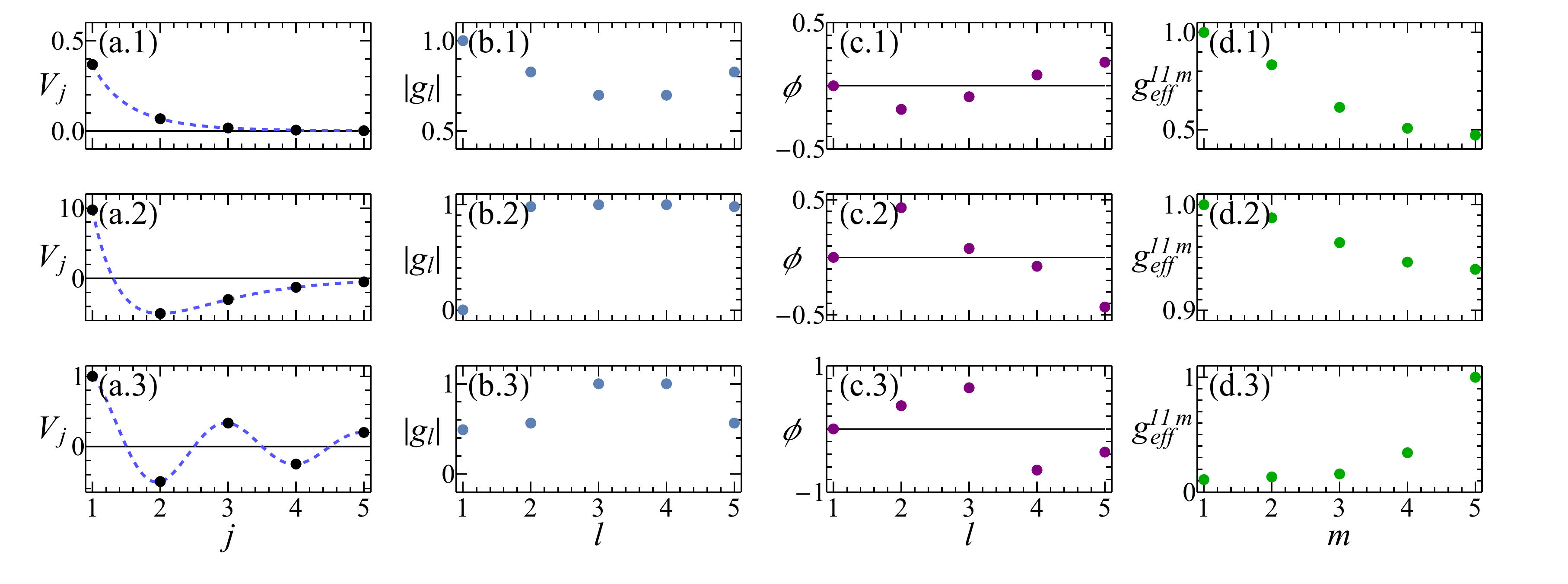}
\captionsetup{width=0.98\textwidth,justification=centerlast,font=small}
\caption{[color on-line] Examples  different synthetic interactions that can be implemented using traveling waves as mode functions. Panel (a) represents the effective synthetic interaction potential $V_j$. We show a Yukawa potential ($V_j=V_j = e^{-j}/j$) in panel (1), a Morse potential in panel (2)$ \left(V_j=5\left[ (1-e^{-(j-2)})^2-1 \right]\right)$ and a Bessel potential in panel (3) ($V_j=\pi y_0(\pi x)$), where $y_0$ is the Bessel function of the second kind.
Panels (b) and (c) show the parameters using a single cavity and five probes described by the effective intensity $|g_l|$ (normalized) and the phase $\phi=\mathrm{Arg}(g_l)$, the effective interaction is $W_{ij}\propto V_i^*V_j$ ($\theta_{c_1}=0$, $\cos \theta_{p_l}=1-2l/5$, $\lambda=2a$). Panel (d)  shows the effective couplings for five cavity modes and one probe corresponding to (a), the effective interaction is $W_{ij}\propto V_{|i-j|}$ ($\theta_{p_1}=0$, $\cos \theta_{c_l}=1-l/5$, $\lambda=2a$). The dashed lines are a guide to eye of the effective mode interaction.} \label{fig:potential}
\end{figure*}
The spatial profile of the interaction described by (\ref{eq:Heff2}) can be tuned changing the light mode functions or/and the intensity and phase of the probe beams. As in the previous examples, we consider traveling waves as mode functions for the light and we note that 
\begin{eqnarray}\HH^D_2&=&\gamma_{\mathrm{eff}}\sum_{l,m=1}^{R} \sum_{i,j}\left(g_l^*g^{\phantom{*}}_m e^{- i (\b{k}_{c_1}-\b{k}_{p_l}) \cdot \b{r}_i}\times
\nonumber\right.\\
 && \left.e^{i (\b{k}_{c_1}-\b{k}_{p_m}) \cdot \b{r}_j}\h{n}_i \h{n}_j +  \mathrm{H.c.}\right)
 \label{eq:modefun1}
\end{eqnarray}
with $\gamma_{\mathrm{eff}}=\Delta_{pc}/(\Delta_{pc}^2+\kappa_c^2)$, as the pumps have the same wavelength $\lambda_{p_l}=\lambda_{p_m}$, such that all the detunings  are the same: $\Delta_{p_lc}=\Delta_{p_mc}=\Delta_{pc}$.  This is equivalent to the product of two Discrete Fourier Transforms (DFT) if $a(\b{k}_{c_1}-\b{k}_{p_l}) \cdot \hat e_{\b{r}_i}=2 \pi l/R$ for all the probe beams $l=1,2,...R$, and $\hat e_{\b{r}_i}$ is the unit vector of $\b{r}_i$ and $a$ the lattice spacing. {Importantly, these conditions fix only the directions of the probe beams ($\cos \theta_{p_l}=\cos \theta_{c_1} - \frac{\lambda}{a} \frac{l}{R}$) and not their intensities or phases, i. e. the coefficients $g_{p}$. }Furthermore, the probe beams divide the optical lattice in $R$ macroscopically occupied spatial modes which interact according to the Hamiltonian,
\begin{align}\label{interaction2}
\HH^D_{2}=\gamma_{\mathrm{eff}}\sum_{i,j=1}^R \left(V_i^* V_j \h{N}_i \h{N}_j + \mathrm{H.c.}\right)
\end{align}
where $V_j= \sum_{m=1}^{R} g_m   e^{ i\frac{ 2 \pi m j}{R} }$ describes the strength of the interaction between the \emph{spatial modes} defined by the light scattering. Tuning the probe intensities and their relative phase, the function $V_j$ can be modified to design to any spatial profile, as illustrated in Fig.~\ref{fig:potential}.  Importantly, Eq.~(\ref{interaction1}) implies that the coupling between the modes $i$ and $j$ is $W_{ij}\propto V_i^* V_j$ and therefore the interaction matrix $W_{ij}$ does not depend on the \emph{distance} between the spatial modes. This is in contrast to the usual solid state physics scenarios where interactions do not depend on the specific position of two particles but only on their distance. This interaction is akin to multicomponent interactions due to different spin projections in a BEC. Here, the interaction coefficients between each light induced mode have an arbitrary coupling that links all the elements in the manifold.  The scheme we propose opens the possibility of studying new classes of interactions and effects not observable in conventional systems. It is worth mentioning that there is only one distance-dependent interaction function that can be realized with this setup: a cosine profile. Specifically, one has $W_{ij}=W_{|i-j|}$ only if $V_j$ is a pure phase (which can be obtained with one probe and one cavity).

{ In the above paragraph, we assumed the light mode functions to be traveling waves which allowed us to give a simple description of the synthetic interactions in terms of DFT. Importantly, the definition of the spatial modes does not rely on this assumption but only on the fact that the coefficients $J_{jj}$ can have the same value on different lattice sites so that atoms in these positions scatter light with the same phase and intensity. For example, it is possible to realize the case $R=2$  by considering standing waves ($u_i(\b{r}) = \cos(\b{k}_i \cdot \b{r })$) crossed at such angles to the lattice that $\b{k}_0\cdot \b{r }$ is equal to $\b{k}_1\cdot \b{r }$ and shifted such that all even sites are positioned at the nodes, so $J_{ii}=1$ for $i$ odd while $J_{ii}=0$ for $i$ even or the $R=3$ case by imposing $\b{k}_{\mathrm{1,0}}\cdot \b{r}=\pi/4$ so that the coefficients $J_{ii}$ are $J_{ii}=[1,\, 1/2,\, 0,\, 1/2,\, 1,\, 1/2,\, 0...]$. If light mode functions are not traveling waves, the  general form of the interaction between the modes follows $V_j= \sum_{m=1}^{R} g_m J_{jj}$ where $J_{jj}$ is not a simple phase factor. Therefore, in order to engineer a given long range interaction it is not possible to use the simple DFT formalism for computing the coefficients $g_m$ but one has to recur to numerical methods.
}

\subsection{Multiple cavity modes and one probe} 
In order to obtain an interaction that depends solely on the distance between the atoms (or the modes), we turn to the case of one classical probe which scatter photons to $R$ light modes. This scheme can be realized using multiple cavities or a multi mode cavity. 
Considering only the events when light is scattered by the atoms from the probe beam to one of the cavities and neglecting the photon scattering between different cavities, we find that the atom-atom interaction follows
\begin{align}\label{eq:Heff3}
\HH^D_3&= \sum_{m=1}^{R} \left(\frac{g^{1m}_\mathrm{eff}}{2} \h{D}_{1m}^\dagger \h{D}_{1m} + \mathrm{H.c.}\right)
\end{align}
where $g^{1m}_{\mathrm{eff}}={\Delta_{1m}|g_1|^2}/(\Delta_{1m}^2+\kappa_{1m}^2)$. Eq. (\ref{eq:Heff3}) is fundamentally different from (\ref{eq:Heff2}) since here only one sum is present and the interaction does not mix $\h{D}$ operators with different indexes. This allows to engineer long-range interactions that depend on the distance between the lattice sites and are analogous to the usual two-body interactions studied in condensed matter systems. We illustrate this by considering traveling waves as mode functions for the light and (\ref{eq:Heff3}) becomes:
\begin{align}\label{eq:interaction3}
\HH^D_{3}&=\sum_{m=1}^{R} \sum_{i,j}\left( g^{1m}_\mathrm{eff} e^{- i (\b{k}_{p_1}-\b{k}_{c_m}) \cdot (\b{r}_i-\b{r}_j)} \h{n}_i \h{n}_j + \mathrm{H.c.}\right)
\end{align}
Fixing the the direction of the wave vectors of the cavity modes so that $a (\b{k}_{p_1}-\b{k}_{c_m}) \cdot\hat e_{\b{r}_i}=\pi l/R$ for all $l=1,2,...R$ (corresponding to the angles fulfilling $\cos \theta_{c_l}=\cos \theta_{p_1} - \frac{\lambda}{2a} \frac{l}{R}$), the light scattering process defines $R$ spatial modes and we can perform a DFT analysis. However, instead we have a Discrete Cosine Transform.  Specifically, Eq.~(\ref{eq:interaction3}) reduces to 
\begin{equation}
\HH^D_3=2\sum_{i,j}V_{|i-j|} \h{N}_i \h{N}_{j},
\end{equation}
where $V_j= \sum_{m=1}^{R}g^{1m}_{\mathrm{eff}}\cos({\frac{ \pi m j}{R} })$. In contrast to the scheme we considered in the previous sections, here the interaction between mode $i$ and mode $j$ depends solely on the distance between the modes ($i-j$) and can be shaped to any $V_j$ profile changing the detunings and the decay coefficients of the cavities. Therefore,   this scheme allows to realize  quantum simulators that are able to mimic long-range interactions with an arbitrary spatial profile, such that $W_{i,j}\propto V_{|i-j|}$.

{As we see, the global (infinitely long-range) light-matter interaction enables to simulate systems with rather short-range and tunable interactions. Moreover, simulating short-range interaction requires just a small number of light modes. Indeed, the price for this is that we do not simulate an original problem of interacting atoms at sites, but replace it by an effective one, simulating the interaction between the global modes. The effective mode Hamiltonians can be an acceptable representation of an otherwise experimentally hard to achieve quantum degenerate system with finite range interaction. As we will show, such a global, but importantly spatially structured interaction, can still compete with intrinsic short-range processes leading to non-trivial phases. As a result, the quantum phases will have properties of systems due to both short-range and global processes, thus directly benefiting from the collective enhancement of the light-matter coupling.}

 In the next sections, we will show the emergent phases that are formed due to particular choices of the light profile used. Also, we will study the general aspects of the effective Hamiltonians by looking at the semi-classical (leading to DOL) and quantum contributions (leading to QOL) in them.

 \section{General decoupling scheme for arbitrary light-field structure: light-induced modes}
 
 In this section we elaborate further on the structure given by the constants that can be designed by the geometry of the system.  As the $J$'s have a certain spatial dependency for a subset of sites,  a sub-lattice structure is imprinted to the atoms. A experimentally relevant particular case, is  when the atoms are located in the diffraction minima of the cavity field (one cavity and one probe, with $R=2$)~\cite{Esslinger2015}. As we have shown in the previous section, the spatial structure of light gives a natural basis to define the atomic modes. The coupling coefficients $J_{i,j}$ can be designed to periodically repeat in space with a certain set of weights. The symmetries broken in the system are inherited from such a periodicity: all atoms equally coupled to light belong to the same mode, while the ones coupled differently belong to different modes. To be precise, in a single cavity with a single mode and pump for the homogeneous scattering in a diffraction maximum, $J_{i,j}=J_B$ and $J_{j,j}=J_D$, one spatial mode is formed. Alternatively, when light is scattered in the main diffraction minimum (at $90^{\circ}$ to the cavity axis), the pattern of light-induced modes alternates sign as in the staggered field, $J_{i,j}=J_{j,i}=(-1)^jJ_B$ and $J_{j,j}=(-1)^jJ_D$. This gives two spatial density modes (odd and even sites) and four bond modes due to the coupling between nearest neighbor coherences~\cite{PRL2015,NJPhys2015}. The density and bond modes can be decoupled by choosing angles such that $J_D=0$ (by shifting the probe with respect to classical lattice thus concentrating light between the sites) or $J_B=0$~\cite{Wojciech}. The Hamiltonian for a single cavity and a single pump is:
\begin{equation}
\HH_{\mathrm{eff}}=\HH^b+\frac{g_{\mathrm{eff}}}{2}\left(\hat F^\dagger\hat F^{\phantom{\dagger}}+\hat F^{\phantom{\dagger}}\hat F^\dagger\right).
\end{equation}
Next, we separate light matter-correlations and dynamical terms in $\hat F^\dagger\hat F$  performing multi mode on-site mean-field.

The $\hat D^\dagger\hat D$ (density coupling) terms can be expanded as
\begin{eqnarray}
\hat D^\dagger\hat D+\hat D\hat D^\dagger&\approx&\sum_{\varphi,\varphi'}(J_{D,\varphi}^*J_{D,\varphi'}^{\phantom{*}}+c.c.)\langle\hat N_{\varphi'}^{\phantom{*}}\rangle(2\hat N_\varphi-\langle\hat N_\varphi^{\phantom{*}}\rangle),
\nonumber
\\
&+&\delta \hat D^\dagger\hat D
\label{DS1}
\\
\label{DS2}
\delta \hat D^\dagger\hat D&=&2\sum_\varphi |J_{D,\varphi}|^2\delta\hat N_\varphi^2,
\\
\delta\hat N_\varphi^2&=&\sum_{i\in\varphi}(\hat n_i-\rho_i)^2,
\end{eqnarray}
where $\langle\hat N_\varphi\rangle=\sum_{i\in\varphi}\rho_i$ is the mean number of atoms in the mode $\varphi$ and $\rho_i=\langle\hat n_i\rangle$ is the mean atom number at site $i$. The first term in Eq. (\ref{DS1}) is due to the dynamical properties of the light field, these terms exhibit non-local coupling between light-induced modes.  The terms in (\ref{DS2}) are the light-matter correlations and contain the effect due to quantum fluctuations the QOL terms.

The $\hat B^\dagger\hat B$ (bond coupling) terms can be expanded as:
\begin{eqnarray}
\hat B^\dagger\hat B+\hat B\hat B^\dagger&\approx&\sum_{\varphi,\varphi'}(J_{B,\varphi}^*J_{B,\varphi'}^{\phantom{*}}+\mathrm{c.c.})\langle\hat S_{\varphi'}^{\phantom{*}}\rangle(2\hat S_\varphi-\langle\hat S_\varphi^{\phantom{*}}\rangle)
\nonumber
\\
&+&\delta\hat B^\dagger\hat B,
\label{BS1}
\\
\delta\hat B^\dagger\hat B&=&2\sum_\varphi |J_{B,\varphi}|^2\delta \hat S_\varphi^2,
\label{BS2}
\end{eqnarray}
\begin{eqnarray}
\delta \hat S_\varphi^2&=&
\sum_{\langle i,j \rangle\in\varphi}(
(\hat b_i^\dagger\hat b_j^{\phantom{\dagger}}+h.c.-\langle\hat b_i^\dagger\hat b_j^{\phantom{\dagger}}+ \mathrm{H.c.}\rangle)^2
\nonumber
\\
&+&
\sum_{\langle i,j,k \rangle\in\varphi}
\big(
\hat b^\dagger_i\hat b^\dagger_k(\hat b^{\phantom{\dagger}}_j)^2
+(\hat b^\dagger_j)^2\hat b^{\phantom{\dagger}}_i\hat b^{\phantom{\dagger}}_k
+
2\hat n_i^{\phantom{\dagger}}\hat b^\dagger_k\hat b^{\phantom{\dagger}}_j
+\hat b^\dagger_k\hat b^{\phantom{\dagger}}_j
\nonumber\\
&&\phantom{\sum_{\langle i,j,k}
}-(\hat b^\dagger_k\hat b^{\phantom{\dagger}}_j+\hat b^\dagger_i\hat b^{\phantom{\dagger}}_k+h.c.)\langle\hat b_i^\dagger\hat b_j^{\phantom{\dagger}}+ \mathrm{H.c.}\rangle\big),
\label{BS2}
\end{eqnarray}
where  $\langle i,j,k \rangle$ refers to $i$,$j$ nearest neighbor and $k$ is a nearest neighbor to the pair $\langle i,j\rangle$. The first term in (\ref{BS1}) is due to the dynamical properties of the light field and (\ref{BS2}) are due to light-matter correlations. These are basically all the possible 4 point correlations and tunneling processes between nearest neighbor, as higher order tunneling processes have much smaller amplitudes. The expectation value of the bond operators reduces to,
 \begin{equation}
 \langle \hat S_{\varphi}\rangle=\sum_{\langle i, j\rangle\in\varphi}(\psi_i^\dagger\psi^{\phantom{\dagger}}_j+\psi_j^\dagger\psi_i^{\phantom{\dagger}})
 ,
 \end{equation}
where $\psi_i=\langle\hat b_i\rangle$  is the SF order parameter corresponding to the site ``$i$". The above is the sum of products of order parameters at nearest neighbor sites in the light-induced mode $\varphi$. 

 In fact for most purposes is enough to consider,
\begin{eqnarray}
\delta \hat S_\varphi^2&\approx&
\sum_{\langle i,j \rangle\in\varphi}
([\hat b_i^\dagger\hat b_j^{\phantom{\dagger}}+ \mathrm{H.c.}]-\langle\hat b_i^\dagger\hat b_j^{\phantom{\dagger}}+ \mathrm{H.c.}\rangle)^2
\\
&=&
\sum_{\langle i,j \rangle\in\varphi}\big[\hat b_i^{2\dagger}\hat b_j^{2\phantom{\dagger}}+\hat b_j^{2\dagger}\hat b_i^{2\phantom{\dagger}}+2\hat n_i\hat n_j+\hat n_i+\hat n_j
\nonumber\\
&-&2(\psi_i^*\psi_j^{\phantom{\dagger}}+\psi_j^*\psi_i^{\phantom{\dagger}})(\hat b_i^\dagger\hat b_j^{\phantom{\dagger}}+\hat b_j^\dagger\hat b_i^{\phantom{\dagger}})+(\psi_i^*\psi_j^{\phantom{\dagger}}+\psi_j^*\psi_i^{\phantom{\dagger}})^2\big],
\nonumber\\
\label{tp}
\end{eqnarray}
as these terms have a more significant effect in the effective Hamiltonian compared to the nearest-neighbor coupling to nearest neighbors (the $\langle i,j,k\rangle$ terms). These are the quantum fluctuations in the SF order parameters. The terms in the first line of Eq.(\ref{tp}) are due to two-particle hole excitations at adjacent sites. These will change the landscape of the supported quantum phases in the system, as they introduce  a mechanism to break translational invariance via a DW instability and the modification to quantum fluctuations. Importantly, the QOL can generate a DW instability. 

Note that for a homogeneous ideal superfluid state ($U=0$),  
\begin{equation}
\sum_{\varphi}\langle\delta \hat S_\varphi^2\rangle=\sum_{\varphi}\langle\hat S_\varphi\rangle=2zN_s|\psi|^2
\end{equation}
where we have used the coordination number (the number of nearest neighbors) is defined as $z=2d$ for a $d$-dimensional square lattice. As we can expect, the fluctuations of the SF order parameter in this limit are { Poissonian}.

The terms that arise from the product of $\hat B$ and $\hat D$ (bond-density coupling) are:
\begin{eqnarray}
\hat B^\dagger\hat D+\hat D\hat B^\dagger+ \mathrm{H.c.}&\approx&2\sum_{\varphi,\varphi'}(J_{B,\varphi}^*J_{D,\varphi'}^{\phantom{*}}+\mathrm{c.c.})
(\langle\hat S_{\varphi'}^{\phantom{*}}\rangle\hat N_\varphi
\nonumber\\
&+&\langle\hat N_\varphi^{\phantom{*}}\rangle\hat S_{\varphi'}^{\phantom{*}}
-\langle\hat S_{\varphi'}^{\phantom{*}}\rangle\langle\hat N_\varphi\rangle))
\nonumber
\\
&+&(
\delta\hat B^\dagger\hat D+\delta\hat D\hat B^\dagger+ \mathrm{H.c.}),
\label{BDS1}
\end{eqnarray}
\begin{eqnarray}
\delta\hat B^\dagger\hat D+\delta\hat D\hat B^\dagger+ \mathrm{H.c.}&=&\sum_\varphi \big[(J_{B,\varphi'}^*J_{D,\varphi'}^{\phantom{*}}+\mathrm{c.c.})\delta \hat C_{\varphi'}
\nonumber\\
&&+ \mathrm{H.c.}\big],
\label{BDS2}
\end{eqnarray}
\begin{equation}
\delta \hat C_{\varphi'}=
\sum_{\langle i,j \rangle\in\varphi'}
(\hat b_i^\dagger\hat b_j^{\phantom{\dagger}}+ \mathrm{H.c.}-
\langle\hat b_i^\dagger\hat b_j^{\phantom{\dagger}}+ \mathrm{H.c.}\rangle)(\hat n_i-\rho_i)
\label{CS1}
\end{equation}
where $\langle\delta \hat C_{\varphi'}\rangle$ 
is the sum of  the ``local" covariances per mode given by the bond operator modes $\varphi'$. The additional terms $\delta\hat B^\dagger\hat B$, $\delta\hat D^\dagger\hat D$, $\delta\hat B^\dagger\hat D$ and $\delta\hat D^\dagger\hat B$ have a local character that alters the system at the quantum level, coupling the local densities to the local tunneling processes.  

Additional terms might be considered in the above expansions and their generalization is straightforward by removing the restriction over the sums beyond nearest neighbor.  It is evident from the decomposition and the expansion that the semi classical terms, given by $\langle \hat F\rangle\hat F^\dagger$ and $\langle \hat F^\dagger\rangle\hat F$ have a non-local (global) character coupling all the illuminated sites and imprinting structure in the interaction.  Light scattering from the atoms can suppress or enhance quantum terms by properly choosing the detuning with respect to the cavity and in addition the light mode structure leads to a combination of novel phases of matter not supported without cavity light. When atoms scatter light maximally, the terms due to quantum fluctuations are strongly smeared out as their behavior scale with  $N_s$ compared with the factor of $N_s^2$ of semi classical terms. This occurs when $g_\mathrm{eff}<0$ and the familiar scenario of self-organized states emerges, however in the case when  $g_\mathrm{eff}>0$ quantum fluctuations become relevant as atoms scatter light minimally and the QOL becomes important. Thus, by suppressing self-organization one has access to the effects due to true quantum fluctuations otherwise not visible.
\subsection{Effective mean-field Hamiltonian components} 
Next, we introduce on site mean-field theory to represent the above terms defining superfluid order parameters per site such that $\langle\hat b_i\rangle=\psi_i$ and we consider for simplicity a square lattice in $d$ dimensions. Separating in short-range contributions, due to quantum fluctuations ($\HH^F_{Q}$) and non-local contributions due to semi-classical terms ($\HH^F_{C}$) we obtain for $\hat F\hat F^\dagger+\hat F^\dagger\hat F=\HH^F_{Q}+\HH^F_{C}$,

\begin{widetext}
\begin{eqnarray}
\HH^F_Q&=&2\sum_i\left[|\gamma_{D,i}|^2(\hat n_i-\rho_i)^2
+
z(\gamma_{D,i}^*\gamma_{B,i}+\mathrm{c.c.})(\hat n_i\hat\beta_i+\hat\beta_i\hat n_i-2\langle\hat\beta_i\rangle\rho_i)
\right.
+z|\gamma_{B,i}|^2(2\hat n_i\hat n_{-i}+\hat n_{i}+\hat n_{-i}
\nonumber\\
&-&
\left.
2\langle\hat\beta_i\rangle\hat\beta_i+\langle\hat\beta_i\rangle^2+\langle\hat b^{\dagger 2}_i\rangle\hat b_{-i}^2+\langle\hat b^{\dagger 2}_{-i}\rangle\hat b_i^2+\langle\hat b^{2}_i\rangle\hat b_{-i}^{\dagger 2}+\langle\hat b^{2}_{-i}\rangle\hat b_i^{\dagger 2}-\langle\hat b^{\dagger 2}_i\rangle\langle\hat b^{2}_{-i}\rangle-\langle\hat b^{\dagger 2}_{-i}\rangle\langle\hat b^{2}_{i}\rangle)
\right],
\\
\HH^F_{C}&=&\sum_i\left[\langle\gamma_{D,i}^*(\hat D+\hat B)+\mathrm{H.c.}\rangle
(2\hat n_i-\rho_i)
+z\langle\gamma_{B,i}^*(\hat D+\hat B)+\mathrm{H.c.}\rangle
(2\hat\beta_i-\langle\hat\beta_i\rangle)\right],
\end{eqnarray}
\end{widetext}
where $\hat \beta_i=\psi_{-i}^*\hat b^{\phantom{\dagger}}_{i}+\psi_{i}^*\hat b_{-i}^{\phantom{\dagger}}+\psi^{\phantom{*}}_{-i}\hat b_{i}^\dagger+\psi^{\phantom{*}}_{i}\hat b_{-i}^\dagger-(\psi_{-i}^*\psi_i^{\phantom{*}}+\mathrm{c.c.})$, $\langle\hat\beta_i\rangle=\psi_{-i}^*\psi_i+\mathrm{c.c.}$, with 
\begin{equation}
\langle\hat D\rangle=\sum_i\gamma_{D,i}\rho_i\quad \textrm{and}\quad\langle\hat B\rangle= z\sum_i\gamma_{B,i}\langle\hat\beta_i\rangle.
\end{equation}
we have used $\gamma_{D,i}=J_{i,i}$ and $\gamma_{B,i}=J_{i,nn(i)}=J_{i,-i}$ where $nn(i)$ is a nearest neighbor of  the site $i$. The sub index $-i$ in operators means nearest neighbor of the site $i$. $\HH^F_{Q}$ are the quantum optical lattice contributions and $\HH^F_C$ are the dynamical contributions to the optical lattice.

In the site decoupling scheme, we have considered that given two operators $\hat x_l$, $\hat y_m$:
$\hat x_l\hat y_m\approx\langle \hat x_l\rangle\hat y_m+\langle\hat y_m\rangle\hat x_l-\langle \hat x_l\rangle\langle \hat y_m\rangle
$. All this provided that we neglect fluctuations between operators at different sites, we do a partial de-correlation. Here 
we assumed that the creation and destruction of a particle at a particular link between nearest neighbors is symmetric, hopping to and from the same link is symmetric. Note that the particular coupling between nearest neighbor sites in the $\hat\beta_i$ operator needs to be constructed consistent with the mode structure given by the light in the effective theory as coupling to and from a particular link in the light mode Hamiltonian might not be symmetric. { We have that for the pair of nearest neighbor sites $i$, $j$: $\langle i, j\rangle$ we have $\langle 1,2\rangle=\langle 2,1\rangle$, but $\langle 1,2\rangle$ is not necessary equal to $\langle 2,3\rangle$ as this symmetry might be broken if the Wannier integrals are $J_{12}\neq J_{23}$. } 

The above mean-field is a good approximation for small local fluctuations in the particle number and the order parameters in each sub-lattice. If the fluctuations between sub-lattices grow, then the mean-field requires modification. The above is accurate as long as: $\Delta(\hat n_{i})\leq 1$, $\Delta(\hat b_{i})\leq 1$, where $\Delta(\hat X_i)^2=\langle \hat X_i^2\rangle- \langle \hat X_i\rangle^2$ (on-site fluctuations). This constraints the description to small amplitude density wave (DW) states and  superfluid (SF) components with small amplitude difference. 

All the physics regarding atom and light depends on the underlying  pattern of the $J$'s  and the spectrum of $\HH^{\mathrm{eff}}$. 

\section{Effective Hamiltonians.} 
The representation of $\HH^F_Q$ and $\HH^F_C$ makes it clear that we can construct an effective mode representation in mean-field approximation for the full $\HH_\mathrm{eff}^b$ depending on the pattern of the $J$'s, such that:
\begin{equation}
\HH^b_{\mathrm{eff}}\approx\HH^b+\frac{g_\mathrm{eff}}{2}(\HH^F_C+\HH^F_Q)
\end{equation}

The effective Hamiltonian considering only density coupling ($\hat D^\dagger\hat D+\hat D\hat D^\dagger$) is
\begin{eqnarray}
\HH^b_{\mathrm{eff}}&=&
\sum_{\varphi}\Big[\sum_{i\in\varphi}\left(-t_0\hat\beta_i+\frac{U_\varphi}{2}\hat n_i(\hat n_i-1)
\right.\nonumber
\\
&-&\left.2g_{\mathrm{eff}}|J_{D,\varphi}|^2\rho_i\hat n_i\right)
-\mu_\varphi\hat N_\varphi-g_{\mathrm{eff}}c_{D,\varphi}\Big],
\nonumber\\
\mu_\varphi&=&\mu-g_{\mathrm{eff}}\eta_{D,\varphi},
\\
U_\varphi&=&U+2g_{\mathrm{eff}}|J_{D,\varphi}|^2,
\end{eqnarray}
with $\eta_{D,\varphi}=\sum_{\varphi'}(J_{D,\varphi}^*J^{\phantom{*}}_{D,\varphi'}+\mathrm{c.c.})\langle \hat N_{\varphi'}\rangle-|J_{D,\varphi}|^2$ and $c_{D,\varphi}=\sum_{\varphi'}(J_{D,\varphi}^*J^{\phantom{*}}_{D,\varphi'}+\mathrm{c.c.})\langle\hat N_{\varphi'}\rangle \langle \hat N_{\varphi}\rangle/2-|J_{D,\varphi}|^2\sum_{i\in\varphi}\rho_i^2$. The many-body interaction $U_\varphi$ and the chemical potential $\mu_\varphi$ inherit the pattern induced by the quantum potential that depends on light-induced mode structure given by $\varphi$.  Thus, each mode component sees a different on site interaction and chemical potential, while there is an additional dependency of the chemical potential on the density. The modification to the on-site interaction is the QOL effect, while the modification to the chemical potential is the DOL effect. As the $\hat \beta_i $ operator couples nearest neighbor sites, in principle one needs 2 on-site modes even for one light induced mode. However, this special case does not break translational symmetry and due to this $\psi_i=\psi_{-i}=\psi$. For more than one light induced mode the number of light induced modes will depend on the number of different values of $J_{D,\varphi}$. 

In the case of only off-diagonal bond scattering  ($\hat B^\dagger\hat B+\hat B\hat B^\dagger$) , we have
\begin{eqnarray}
\HH^b_{\mathrm{eff}}&=&\sum_{\varphi}\Big[-t_\varphi\hat S_\varphi+ g_{\mathrm{eff}}|J_{B,\varphi}|^2\delta \hat S_\varphi^2-g_{\mathrm{eff}}c_{B,\varphi}\Big]
\nonumber\\
&+&\sum_i\left(\frac{U}{2}\hat n_i(\hat n_i-1)-\mu\hat n_i\right),
\\
t_\varphi&=&t_0-g_{\mathrm{eff}}\eta_{B,\varphi},
\end{eqnarray}
with 
$\eta_{B,\varphi}=\sum_{\varphi'}(J_{B,\varphi}^*J^{\phantom{*}}_{B,\varphi'}+\mathrm{c.c.})\langle \hat S_{\varphi'}\rangle$ and $c_{B,\varphi}=\sum_{\varphi'}(J_{B,\varphi}^*J^{\phantom{*}}_{B,\varphi'}+\mathrm{c.c.})\langle\hat S_{\varphi'}\rangle \langle \hat S_{\varphi}\rangle/2$. The effective tunneling amplitude $t_\varphi$ couples the SF components of all the light-induced modes $\varphi$, this is the DOL effect. The terms due to $\delta \hat S_\varphi^2$ induce non-trivial coupling between nearest neighbor sites and lead to the formation of a density wave instability with more than one light-induced mode, this is relevant whenever quantum fluctuations are not smeared out by the semi classical contribution, this is the QOL effect.  

The full Hamiltonian including cross terms products of $\hat D$ and $\hat B$ can be written in the on-site mode dependent decomposition for convenience as,
\begin{widetext}
\begin{eqnarray}
\HH^b_{\mathrm{eff}}&=&
\sum_i\left[-z\left(t_{BD,i}-{g_\mathrm{eff}}(\gamma_{D,i}^*\gamma_{B,i}^{\phantom{*}}+\mathrm{c.c.})\hat n_i\right)\hat\beta_i-\left(\mu_{DB,i}- z g_\mathrm{eff}(\gamma_{B,i}^*\gamma_{D,i}^{\phantom{*}}+\mathrm{c.c.})\hat\beta_i\right)\hat n_i
\right.
\nonumber\\
&+&
zg_{\mathrm{eff}}|\gamma_{B,i}^{\phantom{*}}|^2(2\hat n_i\hat n_{-i}+\hat n_{i}+\hat n_{-i}
+
\langle\hat b^{\dagger 2}_i\rangle\hat b_{-i}^2+\langle\hat b^{\dagger 2}_{-i}\rangle\hat b_i^2+\langle\hat b^{2}_i\rangle\hat b_{-i}^{\dagger 2}+\langle\hat b^{2}_{-i}\rangle\hat b_i^{\dagger 2})
\nonumber\\
&+&
\left.\frac{U_{D,i}}{2}\hat n_i(\hat n_i-1)-c_{DB,i}
\right],
\label{fh}
\end{eqnarray}
with effective non-linear parameters,
\begin{eqnarray}
t_{DB,i}&=&t_0-g_\mathrm{eff}\left(\langle\gamma_{B,i}^*(\hat D+\hat B)+\mathrm{H.c.}\rangle
-2z|\gamma_{B,i}^{\phantom{*}}|^2\langle\hat\beta_i\rangle\right)
\\
\mu_{DB_i}&=&\mu-g_\mathrm{eff}\left(\langle\gamma_{D,i}^*(\hat D+\hat B)+\mathrm{H.c.}\rangle
-2|\gamma_{D,i}^{\phantom{*}}|^2\rho_i-|\gamma_{D,i}|^2\right)
\\
U_{D,i}&=&U+2g_\mathrm{eff}|\gamma_{D,i}|^2
\end{eqnarray}
\begin{eqnarray}
c_{DB,i}&=&
\frac{zg_\mathrm{eff}\langle\hat\beta_i\rangle}{2}\langle\gamma_{B,i}^*(\hat D+\hat B)+\mathrm{H.c}\rangle
+\frac{g_\mathrm{eff}\rho_i}{2}
\left(
\langle\gamma_{D,i}^*(\hat D+\hat B)+\mathrm{H.c.}\rangle
-2|\gamma_{D,i}^{\phantom{*}}|^2\rho_i
\right)
\nonumber\\
&+&\frac{z g_\mathrm{eff}}{2}\left(2(\gamma_{D,i}^*\gamma_{B_i}^{\phantom{*}}+\mathrm{c.c.})\langle\hat\beta_i\rangle\rho_i-2|\gamma_{B,i}^{\phantom{*}}|^2(\langle\hat b^{\dagger 2}_i\rangle\langle\hat b^{2}_{-i}\rangle+\langle\hat b^{\dagger 2}_{-i}\rangle\langle\hat b^{2}_{i}\rangle-\langle\hat\beta_i\rangle^2)\right)
\end{eqnarray}
\end{widetext}
The tunneling amplitudes $t_{DB,i}$ and the chemical potentials  $\mu_{DB,i}$ are renormalized by both semi-classical contributions and due to quantum fluctuations, both DOL and QOL contributions are relevant. The on-site interactions $U_{D,i}$ get modified by quantum fluctuations in the density (QOL effect), while the constants $c_{DB,i}$ are corrections to avoid over counting. The terms in the second line of Eq.(\ref{fh}) contain the effect of the fluctuations in the order parameter (QOL effect). These are an effective nearest neighbor interaction $\hat n_i\hat n_{-i}$, additional chemical potential shifts  and all the two particle-hole excitations between nearest neighbors. In what follows we will show several different effects and quantum many-body phases that occur as one designs the profile for illumination leading to different effective Hamiltonians. 

The key aspect of our approach is to take advantage of the decomposition in the light induced mode basis to generate the effective Hamiltonians and analyze the competing emergent phases. The use of this basis will simplify greatly the estimation of the phase diagram of the system based on effective model Hamiltonians, easy to interpret from their building blocks while retaining enough relevant features to uncover the emergence of unconventional phases of quantum matter.

\section{Homogeneous light scattering}

\subsection{Homogeneous density coupling: superfluid with limited fluctuations}

\begin{figure}[t!]
\begin{center}
\includegraphics[width=0.43\textwidth]{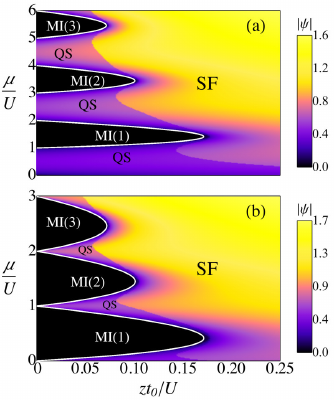}
\captionsetup{width=0.43\textwidth,justification=centerlast,font=small}
\caption{ 
[color on-line] Phase diagram for the Bose-Hubbard model with quantum light (a) and without quantum light (b). SF phases have order parameter $|\psi|\neq0$ 
In the MI($n$) phases the density per site $\rho=n\in\mathbb{Z}_0^+$ and $|\psi|=0$. In (a) SF phases at incommensurate phases appear with limited atomic fluctuations (regions in between MI lobes). In this region (shaded areas) the state of the system is composed of a superposition two Fock states (QS) with the lowest filling for a given density $\rho$, see main text. Parameters are: (a) $J_D=1.0$, $g_{\mathrm{eff}}N_s=0.25U$ (b) $g_{\mathrm{eff}}N_s=0$; $z=6$ (3D) $N_s=100$. The grey scale [color] bar denotes the SF component.}
\label{FH1}
\end{center}
\end{figure}

\begin{figure}[t!]
\begin{center}
\includegraphics[width=0.43\textwidth]{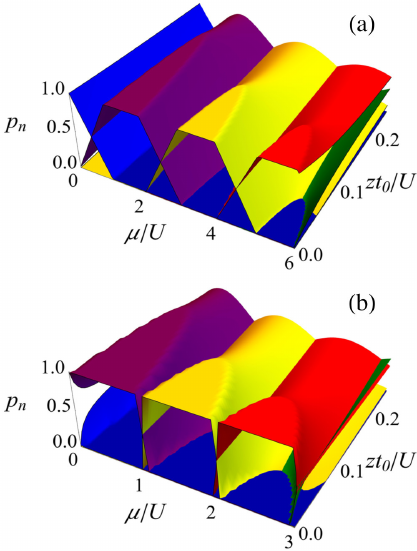}
\captionsetup{width=0.43\textwidth,justification=centerlast,font=small}
\caption{ 
[color on-line] Fock state components probabilities $p_n=|c_n|^2$  of the ground state with quantum light (a) and without quantum light (b). Different shades of grey [colors] correspond to Fock state components: $c_n|n\rangle$. With quantum light when $zt_0/U=0$ the ground state in between MI plateaus (with one Fock state component) is a superposition of two Fock states. The system without quantum light has either one component (MI) or more than 2.  As tunneling is increased more components will contribute to form the ground state. Parameters are: (a) $J_D=1.0$, $g_{\mathrm{eff}}N_s=0.25U$ (b) $g_{\mathrm{eff}}N_s=0$; $z=6$ (3D) $N_s=100$.}
\label{F1b}
\end{center}
\end{figure}

The simplest scenario is the one posed by having the cavity field couplings (Wannier integrals) $J_{i,j}=J_B$ and $J_{i,i}=J_D$ to be site independent, the diffraction \emph{maxima} setting. It follows that with density coupling ($J_D\neq 0$, $J_B=0$) , since we have not broken any spatial symmetry,  it  is a reasonable assumption to consider that the superfluid order parameter is homogeneous over the illuminated area ($\langle\hat b_i\rangle=\psi_i\approx\psi$) and also the density ($\rho_i=\rho$). Then,
\begin{eqnarray}
\HH^{b}_\mathrm{eff}&\approx&\sum_i\Big[-t_0\hat\beta_i-\mu_D(\rho)\hat n_i
+\frac{U_D}{2}\hat n_i(\hat n_i-1)-c_D
\Big]
,
\nonumber\\
\end{eqnarray}
where $\mu_D(\rho)=\mu-g_{\mathrm{eff}}|J_D|^2(2N_s\rho+1)$, $U_D=U+2g_{\mathrm{eff}}|J_D|^2$, $c_D=g_{\mathrm{eff}}|J_D|^2N_s\rho^2$ and $\hat\beta_i=\psi^*\hat b_i^{\phantom{\dagger}}+\psi\hat b_i^\dagger-|\psi|^2$. From the form of the effective Hamiltonian it is clear that one can change the SF-MI transition by the light as this couples to the on-site fluctuations in  $U_D$. The renormalization of the on-site interaction is a byproduct of the quantum fluctuations introduced by the light, a QOL effect. This is easy to see as one plots as a function of the density~\cite{PRL2015}. Moreover, the chemical potential is modified by the self-consistent density field $\rho=\langle\hat n_i\rangle$ thus the phase diagram as a function of the chemical potential changes drastically and SF regions emerge in between the MI lobes compared to the system without cavity light.  This peculiar superfluid state has reduced atomic fluctuations and we call it the Quantum Superposition (QS) state in what follows. This is an effect of the dynamical optical lattice generated via renormalization of the chemical potential and the additional density dependence. { The state was first seen in calculations in~\cite{Morigi} where a similar phase diagram was reported, but its properties and its nature not discussed previously.  In contrast to their work, here we have seen the effect due to quantum fluctuations via the renormalization of the on-site interaction, we will show the structure of the ground state components, while we will discuss the underlying mechanism of the occurrence of QS due to the emergence of an energy gap below. }  The mean field phase diagram where we show the behavior of the superfluid order parameter as a function of tunneling amplitude $t_0\neq0$ and chemical potential, is shown in Fig.\ref{FH1} (a). In Fig.\ref{FH1} (b) we show the behavior of the system without cavity light. The shaded regions in the plots correspond  approximately to the region where the ground state is made of two Fock state components. {The boundary is defined by the numerical threshold of the two component state, the sum of two components, $|c_n|^2+|c_{n+1}|^2\approx0.995$ while all other $c_{\xi}\approx 0\;\forall\xi>n+1$ with $c_{n}$ the Fock state component amplitudes. The Fock state components are shown in Fig.\ref{F1b}.}
 
In the QS state, for certain parameter regimes the steady state becomes gapless even at large U, and the lowest lying excitations dependent on particle filling, as we sill show.  This in turn restricts the steady state to be a sort of topological superfluid, which is just a quantum superposition between the two lowest energy Fock states in mean-field approximation. This is confirmed by exact diagonalization calculations. The two lowest energy Fock states correspond to adjacent integer filling factors; i.e. ``0" and ``1", ``1" and ``2", etc. in between MI lobes. This occurs for incommensurate fillings because the system in the process of minimizing atomic fluctuations cannot achieve the MI state, thus the optimal energetic alternative is a quantum superposition of just two. These phases are a topological superfluid in the sense that there is not a phase transition to these states, but their constrained ground state is very different from the regular superfluid which is made of several Fock states with different filling factors instead. These states appear due to the steady-state degeneracy in the homogeneous system, and the difficulty of the system to ``lock" in a unique value of the density due to incommensuration. Interestingly, these SF states are gapped with respect to adding other excitations as we will see.

 The phase diagrams, have been computed using the Gutzwiller ansatz,
 \begin{equation}
 |\Psi\rangle=\prod_i\sum_{n}c_{n,i}|n\rangle_i,
 \end{equation}
where $c_{n,i}$ are the amplitudes of the Fock state components for each filling $n\in\mathbb{Z}_0^+$, at each site ``$i$".  In this case, as the sites are indistinguishable, so will be the Fock state amplitudes ($c_{n,i}=c_{n}$), leading to an effective single site problem. The value of the Fock state amplitude probabilities $p_n=|c_n|^2$  is shown In Fig \ref{F1b} (a) with cavity light and (b) without cavity light. Looking at that components of the grounds-state $p_n$, we can see that for $zt_0/U=0$, i.e. below the MI plateau for $\mu/U\in\{1,2\}$ the system is in a a superposition of filling ``0" and ``1" . This  has a characteristic ``X" pattern where at $\mu/U=0.5$ the superposition is 50\%. Similar pattern in between the MI plateaus for higher filling emerges, this is to be contrasted in the system without cavity light [Fig. \ref{F1b} (b)] where the behavior is basically absent on the same scale. { Note that even without light, close to the MI lobes,  QS appears as the boundary is estimated by the numerical condition on the sum of two Fock states to be $\approx0.995$.} The even QS states (50\%/50\% superpositions) between adjacent fillings appear at $\mu/U=2 n +1/2$, $n\in\mathbb{Z}_0^+$. The effect of the QS state configuration for strong on-site interaction ($zt_0/U=0$) can be easily seen in the mean occupation $\rho$, where in between the MI plateaus it behaves linearly. The slope of the function depends on the coupling in the large $N_s$ limit as: $\alpha_{D}=U/(2g_{\mathrm{eff}}|J_D|^2 N_s )$. Increasing the tunneling amplitude the system smoothly reaches the regular SF state with more than one Fock state component, thus there is no phase transition, but still the state is rather different in its structure. As the Fock state components are limited, so will be the on-site atomic fluctuations with in this kind of state are always limited, $\Delta(\hat n)\leq 1/4$. { There is no phase transition to this state but a continuos transition without breaking a symmetry, a crossover.}

The density and order parameter have the following simple solutions for this superposition and MI state in the ground state of the effective Hamiltonian, 
\begin{eqnarray}
\psi&=&\sqrt{(n+1)(\rho-n)(1-\rho+n)},
\\
\rho&=& \alpha_{D}\left(\frac{\mu}{U} -n\right), 
\end{eqnarray}
for $n(\alpha_{D}^{-1}+1)\leq {\mu}/{U}\leq \alpha_{D}^{-1}(n+1)+n $,  corresponding to the QS states;
$\psi=0$ and $\rho=n+1$  for $ \alpha_{D}^{-1}(n+1)+n\leq {\mu}/{U}\leq (n+1)(\alpha_{D}^{-1}+1) $
, corresponding to the MI states; with $n\in\mathbb{Z}_0^+$.  As one increases the atom-photon effective interaction QS states emerge in between MI plateaus~\cite {Atoms2015}.
The results are consistent with exact diagonalization simulations, where discrete steps corresponding to the number of sites appear in the ``X" pattern and $\rho$. The discrete steps smooth out as the number of sites is increased. 

The amplitudes of the Fock states in the QS state are  given by:
\begin{eqnarray}
c_{n}&=&\sqrt{1-|c_{n+1}|^2}=\sqrt{1+n-\rho},
\\
c_{n+1}&=&\sqrt{\frac{\mu \alpha_{D}}{U}-n( \alpha_{D}+1)}=\sqrt{\rho-n}, 
\end{eqnarray}
for $n(\alpha_{D}^{-1}+1)\leq {\mu}/{U}\leq \alpha_{D}^{-1}+n(\alpha_{D}^{-1}+1) $,  corresponding to the QS states at incommensurate fillings;
and $c_{n}=1, c_{n+1}=0$, because for $ \alpha_{D}^{-1}+n(\alpha_{D}^{-1}+1)\leq {\mu}/{U}\leq (n+1)(\alpha_{D}^{-1}+1) $, $\rho=n+1$;
this  corresponds to the MI states; with $n=0,1,2,\dots$ and all other $c_{i}=0$. Thus, the steady state can be written as:
\begin{equation}
|\psi_{\mathrm{st}}(\rho)\rangle=c_{[\rho]}|[\rho]\rangle+c_{[\rho]+1}|[\rho]+1\rangle
\end{equation}
where $[\rho]$ is the integer part of $\rho$.
The energy required to add an excitation on top of the ground state is given by:
\begin{eqnarray}
\Delta E_{\mathrm{QS}}(\rho)&=&U c_{[\rho]+1}^2=U(\rho-[\rho]),
\\
\Delta E_{\mathrm{MI}}(\rho)&=&U[\rho+1]+\frac{U[\rho]}{\alpha_{D}}-\mu
\end{eqnarray}
which means that both the MI states and the QS states have an energy gap with respect to states with higher order fillings. This ``gap" closes smoothly as the density increases until the density reaches commensuration and the Mott gap opens. Note that the QS state is itself gapless as it is a superposition of two Fock states, but adding an excitation has a finite cost depending on how many atoms per site there are. In this sense this is a gapped superfluid but different from the spin like systems~\cite{Wen}. Therefore, for  $g_{\mathrm{eff}}\neq 0$, and non-integer filling we are in a QS state. 

\begin{figure*}[ht!]
\begin{center}

\includegraphics[width=0.9\textwidth]{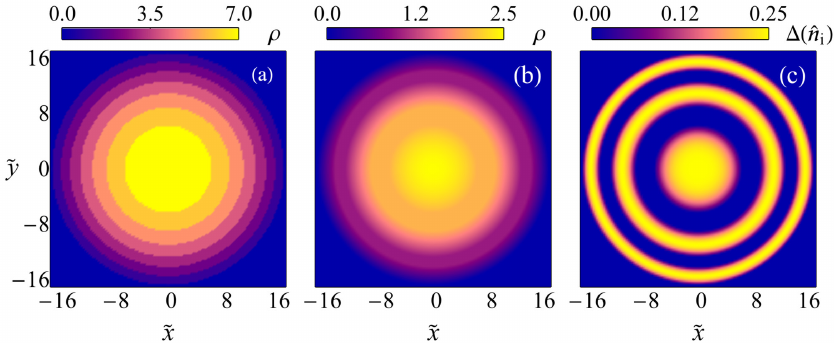}
\captionsetup{width=0.9\textwidth,justification=centerlast,font=small}
\caption{ 
[color on-line] Large interaction limit results ($t_0=0$) with exact diagonalization including harmonic potential in 2D.  Mean atom number per site $\rho$ without cavity light (a) and with cavity light (b), greyscale [color] bar denotes value of the quantity.  (c) Atomic fluctuations $\Delta(\hat n)$, same parameters as in (b) bright regions correspond to the system in the quantum superposition state (QS), dark regions correspond to Mott-Insulator (MI). Atomic fluctuations corresponding to (a) are always zero.  
 Parameters are:  $V=0.025 U /a^2 $,
$g_{\mathrm{eff}}=U/(2N_s)$ (b,c),
$g_{\mathrm{eff}}=0$ (a)
$J_D=1.0$,
$J_B=0$;
$\mu/U=6 (\textrm{a}),7(\textrm{b,c})$. The number of sites is $N_s=10^3$. Here $\tilde x=x/a$, $\tilde y=y/a$ are dimensionless, where $a$ is the lattice spacing. 
}
\label{F1T}
\end{center}
\end{figure*}

 In the case of a very deep classical optical lattice ($t_0=0$), independent of tunneling, the basis of states needed to find the ground state reduces significantly~\cite{Atoms2015}. This allows for an accurate and computational feasible calculation of the large $N_s$ limit, and the exact simulation of harmonic confinement beyond the local density approximation (LDA). In a typical experiment, one has the additional harmonic confinement term given i.e. in 2D by:
\begin{equation}
\HH_T=V\sum_i( x_i^2+y_i^2)\hat n_i
\end{equation}
where $V$ is the harmonic confinement strength. To characterise the state we consider the atomic quantum fluctuations $\Delta(\hat n)$ per site, related to the local compressibility ($\kappa\propto \Delta(\hat n)$). The results including harmonic confinement are shown in Fig.\ref{F1T}. As one would expect from the well known results of the regular Bose-Hubbard model~\cite{Batrouni}, already experimentally seen~\cite{MIShells1,MIShells2}, the MI shell structure emerges without superfluid component if light is not present. However with light in the cavity,  in-between  MI plateaus large and thick QS rings appear.  Surprisingly, now there is a finite SF fraction with minimal fluctuations which the system can support even for a deep lattice. The effect of cavity light can be measured two ways, by looking at the reduction in the number of Mott regions in the system with cavity light and the emergence of finite SF fraction in between Mott regions, previously absent in a deep lattice. 

\subsection{Homogeneous system with cavity assisted tunneling}
In the case where cavity assisted tunneling dominates the behavior of the system ($J_B\neq0$,$J_D=0$), the system requires the inclusion of an additional light induced mode as the coupling between order parameters gets modified. 
\begin{eqnarray}
\HH^b_{\mathrm{eff}}&=&-t_{\varphi_0}\hat S_{\varphi_0}+ g_{\mathrm{eff}}|J_B|^2\delta \hat S_{\varphi_0}^2-g_{\mathrm{eff}}c_{B}
\nonumber\\
&+&\sum_i\left(\frac{U}{2}\hat n_i(\hat n_i-1)-\mu\hat n_i\right),
\\
t_{\varphi_0}&=&t_0-g_{\mathrm{eff}}\eta_{B},
\end{eqnarray}
$\eta_{B}=|J_B|^2\langle \hat S_{\varphi_0}\rangle$ and $c_{B}=|J_B|^2\langle\hat S_{\varphi_0}\rangle^2/2$. The bond operator in mean-field decoupling approximation for this case is given as:  $\hat S_{\varphi_0}\approx z N_s\hat\beta$ with,
$\hat\beta=\psi_{O}^*\hat b^{\phantom{\dagger}}_{E}+\psi_{E}^*\hat b_{O}^{\phantom{\dagger}}+\psi_{O}^{\phantom{*}}\hat b_{E}^\dagger+\psi_{E}^{\phantom{*}}\hat b_{O}^\dagger-(\psi_{O}^*\psi^{\phantom{*}}_E+\mathrm{c.c.})$
 and 
 $\langle\hat\beta\rangle=2\Re(\psi^*_E\psi^{\phantom{*}}_O)$.
The sub-index $O/E$ refer to the light-induced modes between neighboring sites. The terms due to quantum fluctuations are:
\begin{eqnarray}
\delta \hat S_{\varphi_0}^2&\approx&
zN_s(\hat b_E^{2\dagger}\hat b_O^{2\phantom{\dagger}}+\hat b_O^{2\dagger}\hat b_E^{2\phantom{\dagger}}+2\hat n_E\hat n_O+\hat n_E+\hat n_O
\nonumber\\
&-&2(\psi^*_E\psi^{\phantom{*}}_O+\mathrm{c.c.})\hat \beta+(\psi^*_E\psi^{\phantom{*}}_O+\mathrm{c.c.})^2),
\nonumber\\
\label{f2m}
\end{eqnarray}
in general these terms may be omitted as for homogeneous light scattering the term $\eta_B\langle\hat S_{\varphi_0}\rangle\propto N_s^2$. Looking back at the effective Hamiltonian, the quantum fluctuations are an order or magnitude smaller with respect to the number of illuminated sites. As there would be no symmetry broken by these small terms, one can approximate: $\psi\approx\psi_E\approx\psi_O$. Therefore,
\begin{equation}
\HH^b_{\mathrm{eff}}=\left(-z \tilde t_{\varphi_0}\hat\beta
+\frac{U}{2}\hat n(\hat n-1)-\mu\hat n\right)N_s,
\end{equation}
with $\hat\beta=\psi^*\hat b^{\phantom{\dagger}}+\psi\hat b^\dagger-|\psi|^2$, and $\tilde t_{\varphi_0}=t_0-zg_{\mathrm{eff}}N_s|\psi|^2$, where we have omitted the sub-indexes that refer to sites on the mode. The above is an effective single site, single mode Hamiltonian. This effective Hamiltonian is just the regular Bose-Hubbard Hamiltonian in single site decoupling mean-field approximation, where the tunneling amplitude depends self-consistently on the superfluid order parameter via the light matter coupling and the self-consistent constraint: $\psi=\langle\hat b\rangle$. Thus, the physics of the regular BH model get an additional non-linear parametric dependence via the SF component in the system. 

The effect of this is translated to the fact that for $g_{\mathrm{eff}}<0$ the tunneling gets enhanced as the superfluid fraction increments as density grows, suppressing MI phases. For $g_{\mathrm{eff}}>0$, the SF fraction gets suppressed and  the MI lobes grow in size in the phase diagram. The critical renormalized coupling $z \tilde t_{\varphi_0}/U$ for the transition depends on the value of the SF order parameter via $t_{\varphi_0}$. Physically, this can be understood by the fact that the light coupled to the matter at the bonds via the cavity light modifies the tunneling, which changes the mobility of the atoms depending on the detuning. The quantum fluctuations are effectively enhanced ($g_{\mathrm{eff}}<0$) or suppressed ($g_{\mathrm{eff}}>0$). However, in the renormalized effective parameter  $z \tilde t_{\varphi_0}/U$ the phase diagram looks exactly the same as the regular BH model with parameter $z t_0/U$. In essence, in this case, the non-linear parametric dependence is a semi-classical effect that dresses the BH model. 

 It is worth pointing out that, in the case where SF is suppressed ($g_{\mathrm{eff}}>0$) and MI states would occur, we have instead that  for large $g_{\mathrm{eff}}|J_B|^2N_s\gg U$  the terms due to fluctuations of the order parameter ($\delta\hat S_{\varphi_0}^2$) become important. As these terms generate a density wave (DW) instability due to the term $\propto\hat n_E\hat n_O$, then DW order occurs producing a stable ground state with density imbalance $\rho_O\neq\rho_O$, $\rho_{O/E}=\langle\hat n_{O/E}\rangle$. Importantly, this is an intrinsic effect due to the combination of the QOL lattice terms with the other terms in the Hamiltonian.  The system can be supersolid (SS) with $\psi_O\neq 0$ and $\psi_E\neq 0 $ or DW insulator (checkerboard insulator) with  $\psi_O=\psi_E=0$ . As the order parameter gets suppressed, it competes with the fact that it can be energetically favorable to form a DW instead of a MI as part of the energy cost given by the fluctuations gets minimized by the imbalance of density at nearest neighbors instead of a spatially homogeneous configuration. However, mean-field approximation cannot describe accurately large imbalance as it is, as the correlations induced in this limit have been neglected, and more sophisticated methods beyond the scope of this article are needed.

\subsection{Cavity assisted tunneling and density coupling: Stabilization of insulating phases}
In order to consider the limit where both bond coupling ($J_B\neq0$) and density coupling  ($J_D\neq0$) exist, it is necessary to use the full representation given by Eq.(\ref{fh}), since additional mixing of due to the product of $\hat B$ and $\hat D$ occurs. This will translate in density dependent tunneling contributions and density coupling to the order parameters via effective coupling given by the product $J_DJ_B$. As this coupling occurs and due to its  non-linear nature, it is necessary to account for the possibility of mode imbalance.  Therefore, as a consequence of bond coupling, we must consider explicitly two light induced modes ( odd and even sites) in the effective Hamiltonian. Thus, we have,
\begin{widetext}
\begin{eqnarray}
\HH^b_{\mathrm{eff}}&\approx&
\frac{N_s}{2}
\left[
-z\left(t_{BD}-{g_\mathrm{eff}}(J_B^*J_D^{\phantom{*}}+c.c.){(\hat n_E+\hat n_O)}\right)\hat\beta-\left(\mu_{DB}- z g_\mathrm{eff}(J_B^*J_D^{\phantom{*}}+\mathrm{c.c.})\hat\beta\right){(\hat n_E+\hat n_O)}
\right.\nonumber
\\
&+&
2zg_{\mathrm{eff}}|J_B|^2(2\hat n_E\hat n_O+\hat n_E+\hat n_O
+
\langle\hat b^{\dagger 2}_E\rangle\hat b_{O}^2+\langle\hat b^{\dagger 2}_{O}\rangle\hat b_E^2+\langle\hat b^{2}_E\rangle\hat b_{O}^{\dagger 2}+\langle\hat b^{2}_{O}\rangle\hat b_E^{\dagger 2})
\nonumber\\
&+&
\left.
\frac{U_{D}}{2}\hat n_E(\hat n_E-1)+\frac{U_{D}}{2}\hat n_O(\hat n_O-1)-c_{DB}
\right],
\label{h2m}
\\
t_{DB}&=&t_0-g_\mathrm{eff}N_s\left((J_B^*J_D^{\phantom{*}}+\mathrm{c.c.})\rho+z |J_B|^2\left(1-\frac{1}{N_s}\right)\langle\hat\beta\rangle\right)
\\
\mu_{DB}&=&\mu-g_\mathrm{eff}N_s\left(|J_D|^2\left(1-\frac{1}{N_s}\right)\rho+z(J_B^*J_D^{\phantom{*}}+\mathrm{c.c.})\langle\hat \beta\rangle-|J_D|^2\right)
\end{eqnarray}
\begin{eqnarray}
U_{D}&=&U+2g_\mathrm{eff}|J_D|^2
\\
c_{DB,}&=&
{zg_{\mathrm{eff}}N_s\langle\hat\beta\rangle}
\left(
(J_B^*J_D^{\phantom{*}}+\mathrm{c.c.})\rho+z|J_B|^2\langle\hat \beta\rangle
\right)
\nonumber\\
&+&
{g_{\mathrm{eff}}N_s\rho}
\left(|J_D|^2\left(1-\frac{1}{N_s}\right)\rho+z(J_B^*J_D^{\phantom{*}}+J_D^*J_B^{\phantom{*}})\langle\hat \beta\rangle
\right)
\nonumber\\
&+&2 z g_\mathrm{eff}\left((J_B^*J_D^{\phantom{*}}+\mathrm{c.c.})\langle\hat\beta\rangle\rho-|J_B|^2\big[\langle\hat b^{\dagger 2}_E\rangle\langle\hat b^{2}_{O}\rangle+\langle\hat b^{\dagger 2}_{O}\rangle\langle\hat b^{2}_{E}\rangle-(\psi_E^*\psi_O^{\phantom{*}}+\mathrm{c.c.})^2\big]\right)
\label{BD1m}
\end{eqnarray}
\end{widetext}
with $\hat\beta=[\psi_{O}^*\hat b^{\phantom{\dagger}}_{E}+\psi_{E}^*\hat b_{O}^{\phantom{\dagger}}+\psi_{O}^{\phantom{*}}\hat b_{E}^\dagger+\psi_{E}^{\phantom{*}}\hat b_{O}^\dagger+\psi_{O}^*\hat b^{\phantom{\dagger}}_{O}+\psi_{E}^*\hat b_{E}^{\phantom{\dagger}}+\psi_{O}^{\phantom{*}}\hat b_{O}^\dagger+\psi_{E}^{\phantom{*}}\hat b_{E}^\dagger
-(\psi_{O}^*\psi^{\phantom{*}}_E+\mathrm{c.c.})-|\psi_O|^2-|\psi_E|^2]/2$,
 $\langle\hat\beta\rangle=[(\psi^*_E\psi^{\phantom{*}}_O+\mathrm{c.c.})+|\psi_O|^2+|\psi_E|^2]/2$ and $\rho=(\langle\hat n_E\rangle+\langle\hat n_O\rangle)/2$. Where the definition of $\hat\beta$ has been modified to account for the modification of on-site fluctuations by direct coupling of the order parameter of each mode. The results obtained from this mean-field decoupling are qualitatively consistent with simulations via exact diagonalization of small number of sites. A posteriori this can be understood to work because in principle there is no difference for the atoms to know that they belong to a particular light induced mode. Therefore for all purposes the atoms can also see the same light induce mode they belong too while tunneling across the system to a different site with the same probability. As the density couples homogeneously there is no direct imbalance that would break a priori the symmetry.  We estimate the ground state energy, the SF order parameters $\psi_{O/E}$ and the density order parameters $\rho_{O/E}$ using a multi-component Gutzwiller ansatz.  As we have partially de-correlated the Hamiltonian in odd and even modes, we construct an ansatz for the matter state as a product state:
 \begin{equation}
 |\Psi\rangle_b=|\Psi_O\rangle_b\otimes|\Psi_E\rangle_b,\quad 
 \textrm{with}\quad |\Psi_{\xi}\rangle_b=\sum_{n=0}^f \alpha_n^{\xi}|n\rangle_{\xi}
 \end{equation}
with $\xi\in\{O,E\}$ the subspace of the modes Odd or Even. The $\alpha^{O/E}_n$ are the coherent state amplitudes corresponding to the Fock state of filling $n\in\mathbb{Z}^+_0$ of the Hilbert space sector of Odd or Even sites. Afterwards, we generate a functional using that ansatz for the expectation value of the energy of the Hamiltonian given by Eq. (\ref{BD1m}) as a function of the sets of amplitudes $\{\alpha^O\}$ and $\{\alpha^E\}$.  For practical reasons, the number of components of the coherent state superposition ($f$) is truncated depending on the value of the total density or chemical potential and the value of the other parameters of the Hamiltonian. Note that both modes are linked parametrically dependent on the value of the order parameters as these are determined self-consistently via $\psi_{O/E}=\langle\Psi_{O/E}|\hat b|\Psi_{O/E}\rangle$. With this, the solution of the model can be approximately made, provided that imbalance of the order parameters does not involve large fluctuations with respect to their average. The result of this transformation is a global optimization problem in terms of the functional for the expectation values of energy of the system,

 \begin{eqnarray}
E_0=\min_{\{\alpha^O,\alpha^E\}}\langle\HH^b_{\mathrm{eff}}\rangle,\;\textrm{subject to:}
\\
\langle\hat n_O\rangle=\rho_O,\quad\langle\hat n_E\rangle=\rho_E
\\
\langle\hat b_O\rangle=\psi_O,\quad\langle\hat b_E\rangle=\psi_E
\\
\sum_{n=0}^f|\alpha^O_n|^2=1\quad\sum_{n=0}^f|\alpha^E_n|^2=1
 \end{eqnarray} 

\begin{figure}[ht!]
\begin{center}
\includegraphics[width=0.47\textwidth]{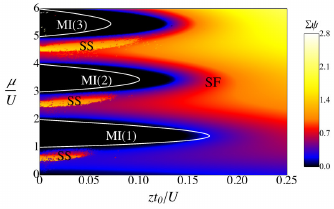}
\captionsetup{width=0.47\textwidth,justification=centerlast,font=small}
\caption{ 
[color on-line] Phase diagram of the system with both density and bond coupling with homogeneous illumination. Mott lobes are stabilized by bond coupling and SS phase appears for incommensurate fillings. The system is in the SF state when the total order parameter $\Sigma\psi=(|\psi_O|+|\psi_E|)/2\neq0$. For MI states $\Sigma\psi=0$. SS appears when $\rho_O\neq\rho_E$ and $\Sigma\psi\neq0$. White lines correspond to the MI-SF transition without light (shifted in chemical potential). Parameters are: $J_D=1.0$, $J_B=0.05$, $g_{\mathrm{eff}}N_s/U=0.25$, $N_s=100$, $z=6$ (3D). The value of the $\Sigma\psi$ component corresponds to the greyscale [color] bar in the SF region.   
}
\label{PDBD2}
\end{center}
\end{figure}

 Furthermore, it is useful to find a decomposition such that: $\HH^b_{\mathrm{eff}}=\HH_{\mathrm{eff}}^O+\HH_{\mathrm{eff}}^E$ and numerically optimize simultaneously both energy contributions.  Following this, we calculate the order parameters and construct  the phase diagram of the model [Fig. \ref{PDBD2} ]. 

Surprisingly, we find that the action of the bond coupling is to stabilize the MI regions in the system as the density increases. Similarly to the case without bond coupling, we find that QS states form in between MI lobes. In the effective Hamiltonian Eq. (\ref{h2m}), the effective tunneling amplitude gets suppressed via the density coupling. Therefore as the number of particles increases per site, for commensurate fillings, the atoms loose mobility stabilizing the MI lobes instead of shrinking the critical tunneling amplitude, as it happens in the case without light (see white lines in Fig.\ref{PDBD2}). This effect occurs as quantum fluctuations get further suppressed via bond coupling as we increase $g_{\mathrm{eff}}$ for large on-site interactions. 
 
 In addition, we find that for incommensurate fillings, it is possible to find regions where in addition to QS states, density imbalance between modes is favorable. Therefore, for large on-site interaction the SS phases can be the competing ground-state. In this case, a  density wave stabilizes instead of a homogeneous state. This can be traced back to the fact that as the order parameters couple additionally to the density via on-site covariances and that densities of each mode couple to both SF order parameters, see Eq. (\ref{h2m}).This is a combination of both the quantum and dynamical terms in the Hamiltonian. Importantly here, both the QOL and DOL combine to generate the density instability while further suppressing fluctuations.  For fixed $g_{\mathrm{eff}}N_s/U$, as $t_0/U$ decreases and the incommensuration occurs, it is favorable for the  SF order parameters to be different. The system moves from the regular SF state the QS state and as the MI boundary is approached a DW can form. This occurs as the non-linear coupling to the densities of each mode compensate for the effect of the on-site interaction on each mode in the energy. Thus, the competition between the quantum fluctuations induced by light and the on-site interaction with the effect of incommensuration trigger this process. For very large $g_{\mathrm{eff}}N_s/U$ close to half-integer filling we can expect that SS phases are further stabilized, and then higher order correlations will become important as density imbalance grows. Then the terms due to the quantum fluctuations of the order parameter will also be significant. This limit is beyond the current mean-field description confined to small fluctuations  of  atom numbers and tunneling processes.  
 
\section{Diffraction minima: Light scattering at $90^\circ$}
\subsection{Emergent super-solids with density coupling}

The possibility to arbitrarily construct sub-lattice structures by carefully choosing the spatial profile of the cavity field pumped to the system allows for the study of complex and more exotic phases. Since now we can break the spatial symmetry along the system by design, this gives rise to the possibility of generating macroscopic phases with space modulation. These emergent phases due to the quantum nature of light and its long range character motivate the study of the possibility of formation of both density wave order (DW) and its combination with superfluid order, the super-solid phase (SS) in a different form. In general, the formation of super-solid order is a long standing elusive question regarding experimental realisation in systems where a lattice Hamiltonian is the adequate description~\cite{DasSarma,Pupillo,RevSS}. Our system differs from the typical Hamiltonians studied because the interaction is infinite range, in the usual studies of the extended Hubbard model~\cite{EBHZoller,EBHLewenstein}, the interaction is typically considered up to nearest neighbors. In our system the cavity back-action couples all sites and this is a fundamental aspect in the modelling of the system.. This system has been recently realized in experiments~\cite{Esslinger2015,Hemmerich2015}, where due to the DOL seen by the atoms both DW and SS can occur.

As we will see the scenario regarding the formation of SS and DW phases is  confirmed and re-shaped. In the case of diffraction minima we have the following pattern of the $J_{D,\varphi}$, in terms of the $\gamma$'s,
$
\gamma_D(i)=(-1)^{i+1}
$.
This generates a double sub-lattice structure, where it is convenient to define lattices $O$ and $E$ corresponding to the positive and negative case for the $\gamma$'s. This configuration can be achieved for a transverse pump in the cavity, pumping light at $90^\circ$ with respect to the cavity axis on the side. The light induced effective structured interaction induces two modes for odd $(O)$ and even $(E)$ sites across the square classical optical lattice. We have then,
\begin{eqnarray}
\hat D&=&J_D\sum_\nu(\hat n_{O,\nu}-\hat n_{E,\nu}),
\end{eqnarray}
and the effective Hamiltonian is:
\begin{equation}
\HH^b_{\mathrm{eff}}=\HH^b+g_{\mathrm{eff}}|J_D|^2\left[\sum_\nu(\hat n_{O,\nu}-\hat n_{E,\nu})\right]^2
\end{equation}
where the sum over $\nu$ goes over $N_s/2$ sites. The effective mean-field Hamiltonian following the general mean-field decoupling scheme is then,
 \begin{eqnarray}
\HH^b_{\mathrm{eff}}&\approx&\HH^O_{\mathrm{eff}}+\HH^E_{\mathrm{eff}}
\\
\HH^\xi_{\mathrm{eff}}&=&\frac{N_s}{2}\Big[-z t_0\hat\beta-\mu_{\xi}\hat n_{\xi}
 +\frac{U_{\xi}}{2}\hat n_{\xi}(\hat n_{\xi}-1)\nonumber\\
&-&g_{\mathrm{eff}}|J_D|^2\rho_{\xi}\hat n_{\xi}-g_{\mathrm{eff}}c_{D,\xi}\Big],
\\
\mu_{O/E}&=&\mu\pm 2g_{\mathrm{eff}}N_s|J_D|^2\Delta\rho,
\\
U_{O/E}&=&U+2g_{\mathrm{eff}}|J_D|^2,
\end{eqnarray}
where $\xi=O/E$ with $c_{D,O/E}=\pm N_s|J_D|^2\Delta\rho\rho_{O/E}/2-|J_D|^2\rho_{O/E}^2/2$, $\hat\beta=\psi_{O}^*\hat b^{\phantom{\dagger}}_{E}+\psi_{E}^*\hat b_{O}^{\phantom{\dagger}}+\psi_{O}^{\phantom{*}}\hat b_{E}^\dagger+\psi_{E}^{\phantom{*}}\hat b_{O}^\dagger
-(\psi_{O}^*\psi^{\phantom{*}}_E+c.c.)$,
 and $\langle\hat\beta\rangle=(\psi^*_E\psi^{\phantom{*}}_O+\mathrm{c.c.})$. It is useful to define $\Delta\rho=(\rho_O-\rho_E)/2$ the emergent DW order parameter and the density $\rho=(\rho_O+\rho_E)/2$. As before, $\langle\hat n_{O/E}\rangle=\rho_{O/E}$ and $\langle\hat b_{O/E}\rangle=\psi_{O/E}$ are self-consistent constraints. We have assumed that these self-consistent parameters are homogeneous in each sub-lattice $O/E$. It is useful to regroup the $\hat\beta$ for the operators of each mode as:
 \begin{equation}
 \hat\beta_{O/E}=2\big(\psi_{E/O}^*\hat b^{\phantom{\dagger}}_{O/E}+\psi_{E/O}^{\phantom{*}}\hat b_{O/E}^\dagger\big)
-(\psi_{O}^*\psi^{\phantom{*}}_E+\mathrm{c.c.})
 \end{equation}
 so that then, the operator part of each sub-lattice Hamiltonian acts on its own sub-lattice Hilbert space, as $2\hat\beta=\hat\beta_O+\hat\beta_E$. Thus, the problem can be cast again in terms of the global optimization problem to find the ground-state using the following expectation values for the energy introducing the Gutzwiller ansatz  for each mode,
 \begin{eqnarray}
 E_0&=&\min_{\{\alpha^O,\alpha^E\}}\big(\langle\HH^O_{\mathrm{eff}}\rangle+\langle\HH^E_{\mathrm{eff}}\rangle\big).
 \end{eqnarray}
 \begin{figure}[t]
\begin{center}
\includegraphics[width=0.43\textwidth]{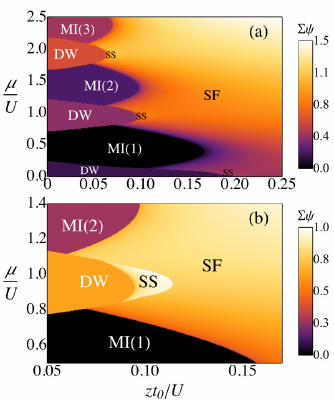}
\captionsetup{width=0.43\textwidth,justification=centerlast,font=small}
\caption{ 
[color on-line] (a) Phase diagram of two component density ordered states for $g_{\mathrm{eff}}<0$, when geometry of the light pumped into the system such that 2 spatial modes can occur  ($R=2$). The system is SF, without any spatial pattern as the interaction is decreased. The SF phase has total order parameter $\Sigma\psi=(|\psi_O|+|\psi_E|)/2\neq 0$ and $\psi_O=\psi_E$.  MI($n$) lobes appear for commensurate densities $n$, $\Sigma\psi=0$, $\rho=(\rho_O+\rho_E)/2=n\in\mathbb{Z}^+_0$ with $\rho_O=\rho_E$. In between MI lobes, DW insulators form with $\Delta\rho=(\rho_O-\rho_E)/2\neq0$, maximal light scattering occurs then and $\Sigma\psi=0$. SS phases ($\Sigma\psi\neq 0$, $\Delta\rho\neq0$ ) are indicated near the boundary between DW and SF states. (b) Phase diagram closer to the DW-SS-SF transition at the tip of the DW insulators in between MI(1) and MI(2). The transition can have an intermediate SS state towards the homogeneous SF phase. Parameters are for (a) and (b): $J_D=1.0$, $J_B=0.0$, $g_{\mathrm{eff}}N_s/U=-0.5$, $N_s=100$, $z=6$ (3D). The grey scale [color] bar denotes $\Sigma\psi$ in the SF region. Quantum phases different from SF (labeled) are denoted by regions with different shades of grey [colors].}
\label{PD2mD}
\end{center}
\end{figure}
 In addition to the self-consistent constraints for the order parameters in the optimization problem, we have: $\sum_{n=0}^f|\alpha_n^{\xi}|^2=1$, using the coherent state expansion for the Gutzwiller ansatz of each mode as in the previous section.
 The quantity $\Delta\rho$ measures the formation of density wave order  in the system in the stationary state.  Density wave order will be present in the system given that $\Delta\rho\neq 0$, this induces a checkerboard pattern in the density over the entire lattice. One can see from the above, that depending on the balance between the different couplings $\mu_{O/E}$ and the original parameters of the Bose-Hubbard model in the absence of quantum light $\HH^b(t_0,\mu,U)$, there exists the possibility for the system to be in different macroscopic phases in the steady state additional to the Mott-Insulator (MI) phases ($\Delta\rho=0$ and $|\psi_{O/E}|=0$) and the superfluid (SF) phases ($\psi_{O}=\psi_E\neq 0$. The system can be additionally to the regular SF an MI phases in a density wave (DW) insulating phase ($\Delta\rho\neq 0$ and $|\psi_{O/E}|=0$) or in a super-solid phase ($\Delta\rho\neq0$ and $|\psi_{O/E}|\neq0$).   Essentially, the  components $\HH_{\mathrm{eff}}^b$ are just  Bose-Hubbard models for the sub-lattices $O/E$ coupled to each other via chemical potentials $\mu_{O/E}$.  The long range effect of the cavity field is encoded in the dependency  between sub-lattices in the self consistent parameters for the mean atom number per site and the superfluid order parameters. These terms due to the effective coupling between sub-lattices induce long range order in our model, diagonal in $\rho_{O/E}$ and off-diagonal in $\psi_{O/E}$. 
 \begin{figure}[th!]
\begin{center}
\includegraphics[width=0.43\textwidth]{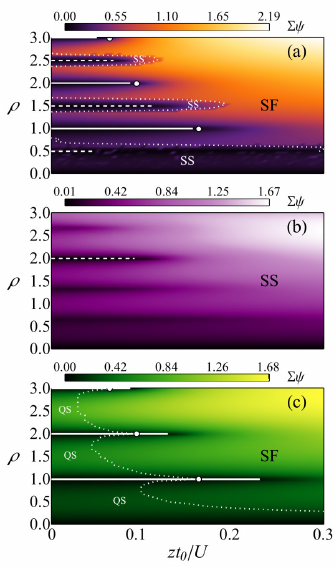}
\captionsetup{width=0.43\textwidth,justification=centerlast,font=small}
\caption{ 
[color on-line] Phase diagrams at fixed density with quantum light for 2 light induced modes ($R=2$). (a) The SF phase has total order parameter $\Sigma\psi=(|\psi_O|+|\psi_E|)/2\neq 0$ and $\psi_O=\psi_E$.  MI($n$) lobes appear for commensurate densities $n$, $\Sigma\psi=0$, $\rho=(\rho_O+\rho_E)/2=n\in\mathbb{Z}^+_0$ with $\rho_O=\rho_E$, white lines. In between MI lobes, DW insulators form with $\Delta\rho=(\rho_O-\rho_E)/2\neq0$, maximal light scattering occurs then and $\Sigma\psi=0$, white dashed lines. SS phases ($\Sigma\psi\neq 0$, $\Delta\rho\neq0$ ) are indicated near the boundary between DW and SF states. (b) Large  $g_{\mathrm{eff}}<0$ regime. Only SS and DW exist (white dashed line). (c)  Large $g_{\mathrm{eff}}>0$ regime. MI insulators (white lines), regular SF and QS (two Fock component SF) exist. The two components in the system are the same. Dashed lines shows the boundary below where QS occurs. White points denote the SF-MI transition point without cavity light. Parameters are for (a)  $g_{\mathrm{eff}}N_s=-0.5U$, (b) $g_{\mathrm{eff}}N_s=-1.25U$ (c) $g_{\mathrm{eff}}N_s=10U$;  $J_D=1.0$, $J_B=0.0$, $N_s=100$, $z=6$ (3D).  The grey scale [color] bar denotes $\Sigma\psi$ in the SF region. }
\label{FD2mDF}
\end{center}
\end{figure}

 Recently similar extended Bose-Hubbard models  where only nearest-neighbor interaction have been considered ~\cite{Miyashita, Ishkin} and in the context of Rydberg interactions~\cite{Pupillo, EBHZoller,EBHLewenstein} finite range models are also actively studied. Additionally, in our effective models, quantum fluctuations in each sub-lattice are modified by the light-matter interaction modifying the Hubbard $U$. The phase diagram of the system is shown in Fig. \ref{PD2mD} as a function of the chemical potential and the effective tunneling amplitude $zt_0/U$; as a function of the density it is shown in Fig.\ref{FD2mDF} (a).  When light scatters maximally $g_{\mathrm{eff}}<0$ the modification due to the quantum fluctuations can be safely neglected as their contribution is strongly smeared out and we have  a DOL (the contribution goes like $N_s$ vs. $N_s^2$ in contrast to the semi classical contribution).  Depending on $g_{\mathrm{eff}}$ with respect to the on-site interaction $U$ there is the formation of DW  lobes in between the typical MI lobes in the system, at half integer fillings. In between the Mott regions as $U$ decreases at fixed $\mu/U$,  we find that SS phases can appear as  intermediate states from the DW towards the SF state as $U$ decreases. The size of the SS and DW phases is strongly influenced by the ratio $|g_{\mathrm{eff}}| N_s/U$.
This is similar to the case where nearest neighbors interaction is considered only in an extended Bose-Hubbard model in addition to a soft-core (finite $U$) with on-site interaction~\cite{Miyashita}, however here the coupling between the sub-lattices is via their difference in mean occupation (the DW order parameter). It is well known that the combination of soft-core bosons and nearest neighbors interactions stabilizes the SS phase against phase separation, we expect our system to  behave likewise. The number of photons scattered is  $\langle\hat a^\dagger\hat a\rangle\propto\Delta\rho^2N_s^2$. Thus, when DW order occurs we expect a large signal in the detector as photons escape the cavity.

In general for $g_{\mathrm{eff}}<0$, when $|g_{\mathrm{eff}}||J_D|^2N_s\gg U$ the system can support only DW and SS phases as then maximum amplitude DW order takes over when $|g_{\mathrm{eff}}||J_D|^2N_s>U/2$. This occurs as the light induced interaction being effectively attractive for one of the modes in the system is equal or stronger than the repulsive on-site interaction, as this occurs, a maximally imbalance state is favoured. Thus the system will reach a maximally imbalanced configuration where either odd or even sites will be empty on average. If interactions are strong enough completely suppressing fluctuations then the atoms will form a checkerboard insulator or otherwise the system will be in the SS state, see Fig.\ref{FD2mDF}(b).The case of a checkerboard insulator for filling 2 has been discussed in~\cite{Reza}, { other recent calculations have also been performed~\cite{Chin,Doner}, consistent with our results}. In principle when $g_\mathrm{eff}$ increases,  additional correlations and strong density imbalance arises and the effective mean-field theory based on the decoupling approximation becomes inaccurate. However qualitative agreement can be found if the total density per site is fixed~\cite{PRL2015}.

When $g_{\mathrm{eff}}>0$, terms due to quantum fluctuations cannot be ignored. The QOL generated  can shift the MI-SF transition depending on the strength of the light pumped into the system as they effectively couple directly to the on-site density fluctuations. Thus light induced atomic quantum fluctuations add to the effect of on-site repulsive interactions shifting the critical point, see Fig.\ref{FD2mDF}(c). If on-site interactions would be completely suppressed still a MI phase could be achieved by means of only cavity light, provided the light matter coupling is strong enough even in a shallow lattice. As light is scattered minimally for $g_{\mathrm{eff}}>0$ the light amplitude directly measures the quantum fluctuations of the matter field, near a MI states the light signal would be strongly suppressed then. In this case the number of photons scattered is  $\langle\hat a^\dagger\hat a\rangle\propto N_s[\Delta(\hat n_O)^2+\Delta(\hat n_E)^2]/2=N_s\Delta(\hat n)^2$.
 
We note, that our mean-field simulations are an  indicative picture of the qualitative form of the phase diagram. Certainly, more sophisticated simulations and experiments could shed more light on the peculiarities of the qualitative behavior depicted in connection with the SS phase and the formation of the DW. 
 
\subsection{Emergent bond order: Phase structured steady-states}
\begin{figure*}[ht!]
\begin{center}
\includegraphics[width=0.95\textwidth]{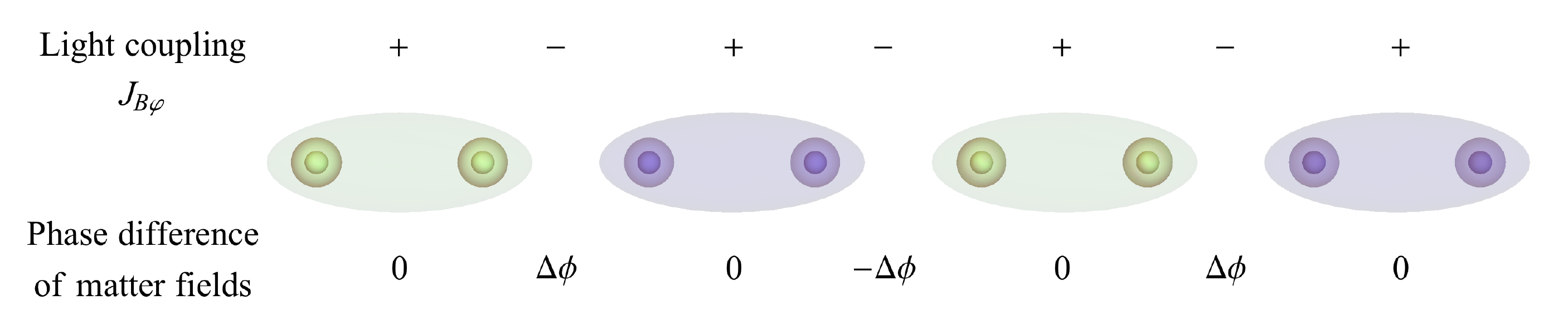}
\captionsetup{width=0.95\textwidth,justification=centerlast,font=small}
\caption{ 
[color on-line] Spatial structure of dimers (ellipsoids) and light imposed phase difference. For $U=0$ atom population in different dimers remains the same. For $U\neq0$ dimers can have different population between them, different shades of grey [colors].}
\label{pattdim}
\end{center}
\end{figure*}

Carefully choosing the light pumped into the system one can achieve maximal coupling of the inter-site density terms (bond coupling) while completely suppressing the density coupling contribution. This can be achieved by illuminating at $90^\circ$ with respect to the cavity axis while shifting the phase of the quantum potential with respect to the optical lattice by $\lambda/4$.  More precisely, the effective light matter coupling depends on the light mode functions which can be chosen such that the nodes of the pumped light field occur at the density maxima of the classical optical lattice. { In 1D, one can choose $u_c({\bf r})=\mathrm{constant}$ and $u_p({\bf r})=\sin(\pi{\bf r}/d)$ with $d$ the lattice spacing. Thus the light couples to the inter-site densities, maximizing $\hat B$  and suppressing entirely $\hat D$ ($J_D=0$) \cite{Wojciech}. Therefore, one obtains a patterned bond interaction where $J_{B,\varphi}=\pm J_B$. Typically $J_B\approx 0.05$ for a classical optical lattice with potential amplitude $V_0\approx5 E_R$ in 1D.} As this occurs, the light couples the matter by imprinting the phase pattern of the light onto the matter. The effective Hamiltonian of the system is then,

\begin{eqnarray}
\HH^b_{\mathrm{eff}}&=&\sum_{\xi=1}^4\HH^\xi_{\mathrm{eff}}
\\
\HH^\xi_{\mathrm{eff}}&\approx&\frac{N_s}{4}\big[\frac{z}{2} t_{\varphi_\xi}\hat\beta_\xi-\mu\hat n_\xi+g_\mathrm{eff}|J_B|^2\delta S_{\varphi_\xi}^2
\nonumber\\
&+&
U \hat n_\xi(\hat n_\xi-1) -g_{\mathrm{eff}}N_s|J_B|^2\tilde c_{B,{\varphi_\xi}}
\big]
\\
t_{\varphi_\xi}&=&t_0-g_{\mathrm{eff}}N_s|J_B|^2\tilde\eta_{B,{\varphi_\xi}},
\\
\hat\beta_\xi&=&
\big[
\psi_{\xi}^*\hat b^{\phantom{\dagger}}_{\xi+1}
+\psi_{\xi+1}^*\hat b_{\xi}^{\phantom{\dagger}}
+\psi_{\xi+1}^{\phantom{*}}\hat b_{\xi}^\dagger
+\psi_{\xi}^{\phantom{*}}\hat b_{\xi+1}^\dagger
\nonumber
\\
&-&
(\psi_{\xi}^*\psi^{\phantom{*}}_{\xi+1}
+\mathrm{c.c.})
\big]
\end{eqnarray}
\begin{eqnarray}
\tilde\eta_{B,\varphi_\xi}&=&\frac{z(-1)^{\xi+1}
}{8}\sum_{\xi'=1}^4(-1)^{\xi'+1}(\psi_{\xi'}^*\psi^{\phantom{*}}_{\xi'+1}+\mathrm{c.c.}),
\nonumber\\
\\
\tilde c_{B,{\varphi_\xi}}&=&\frac{z}{4}(\psi_{\xi}^*\psi^{\phantom{*}}_{\xi+1}+\mathrm{c.c.})\tilde{\eta}_{B,{\varphi_\xi}}
\end{eqnarray}
\begin{eqnarray}
\delta S_{\varphi_\xi}^2&=&\frac{z}{2}\big[\hat b_\xi^{2\dagger}\hat b_{\xi+1}^{2\phantom{\dagger}}+\hat b_{\xi+1}^{2\dagger}\hat b_{\xi}^{2\phantom{\dagger}}
+2\hat n_{\xi}\hat n_{\xi+1}+\hat n_{\xi}+\hat n_{\xi+1}
\nonumber\\
&+&\hat b_\xi^{2\dagger}\hat b_{\xi-1}^{2\phantom{\dagger}}+\hat b_{\xi-1}^{2\dagger}\hat b_{\xi}^{2\phantom{\dagger}}
+2\hat n_{\xi}\hat n_{\xi-1}+\hat n_{\xi}+\hat n_{\xi-1}
\nonumber\\
&-&2(\psi^*_\xi\psi^{\phantom{*}}_{\xi+1}+\mathrm{c.c.})\hat \beta_\xi+(\psi^*_{\xi}\psi^{\phantom{*}}_{\xi+1}+\mathrm{c.c.})^2
\nonumber\\
&-&2(\psi^*_\xi\psi^{\phantom{*}}_{\xi-1}+\mathrm{c.c.})\hat \beta_\xi+(\psi^*_{\xi}\psi^{\phantom{*}}_{\xi-1}+\mathrm{c.c.})^2
\big],
\end{eqnarray}
 where the component $\xi+4$ is the same as $\xi$, while $\langle\hat \beta_\xi\rangle=(\psi_{\xi}^*\psi^{\phantom{*}}_{\xi+1}
 +\mathrm{c.c.})$ and as usual $\langle\hat b_\xi\rangle=\psi_\xi$ defines the order parameters of the light induced modes.
 
  Due to the fact that sites couple differently with period 4, one needs to introduce effectively 4 modes to represent the action of the light-matter interaction in the effective Hamiltonian. In addition to the amplitudes of the order parameters for the 4 modes, now the phase of the matter fields is extremely important. For $g_{\mathrm{eff}}<0$, to compensate for the light induced phase pattern, the system minimizes its energy by maximizing light scattering. Thus, the phase of the matter fields acquires spatial modulation. The self-organization DOL typical mechanism generates this effect. The phase difference between SF order parameters in adjacent sites can be $0$ or $\pm\Delta\phi$ with $\Delta\phi\neq0$. When the phase difference between SF order parameters is zero then a dimer structure forms, as the SF component is spatially indistinguishable in these two adjacent sites. Then, a phase jump in the  order parameters SF occurs generating a distinguishable feature for the next dimer, the pattern then extends across all the lattice in all dimensions. In 1D the pattern is shown in Fig.\ref{pattdim}. In this limit, one can neglect the terms that arise from $\delta S_{\varphi_\xi}^2$, as the modification in the tunneling amplitudes is an order of $N_s$ larger. The determination of the difference in phase, and order parameters can be done in terms of another global optimization problem introducing the Gutzwiller ansatz  for each mode which is now,  
 \begin{eqnarray}
 E_0&=&\min_{\{\alpha^1,\alpha^2,\alpha^3,\alpha^4\}}\sum_{\xi=1}^4\langle\HH^\xi_{\mathrm{eff}}\rangle.
 \end{eqnarray}
 
\begin{figure}[ht!]
\begin{center}
\includegraphics[width=0.43\textwidth]{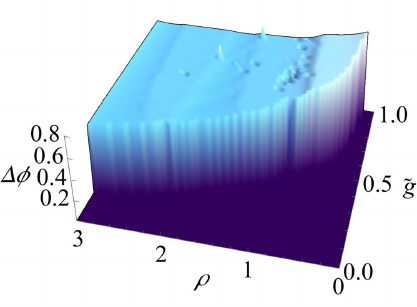}
\captionsetup{width=0.43\textwidth,justification=centerlast,font=small}
\caption{ 
[color on-line] Phase difference $\Delta\phi$ between dimers  for maximal light scattering through the bonds ($J_B\neq0$, $g_{\mathrm{eff}}<0$) as a function of the average density per site $\rho$ and the effective light matter coupling $\tilde g=-2 z g_{\mathrm{eff}}|J_B|^2N_s/t_0$. Whenever $\Delta\phi\neq 0$ the system is in the superfluid dimer phase, otherwise the system is in the normal SF state, $U=0$. { In all simulations the structured ground state has energy with several $E_R$ lower than the homogenous ground state, substructure arises due to numerics.}
}
\label{NID}
\end{center}
\end{figure}

 As a consequence of the particular effective light-matter interaction, for $g_{\mathrm{eff}}<0$, the matter fields self-organize to maximize light scattering, leading to have a well defined spatial pattern in the phase of the matter-wave, akin to the checkerboard patter in the density. Thus, pairs of nearest neighbor phase correlated atoms posses a phase difference that is modulated. We call this state: the superfluid dimer (SFD) state. Parallelism can be drawn concerning the properties of this state with the well known phase modulated state of superconducting fermions, the FFLO/LOFF state~\cite{FFLO1,FFLO2}. {In the FFLO/LOFF state the superconducting order parameter varies spatially periodically akin to the spatial phase variation seen in the dimer state. This can be traced back to the finite momentum transfer induced by light to the atoms via the bond coupling (Addressing $\hat B$). This is similar to the finite momentum acquired by Fermi surface component mismatch in the fermionic system, forming finite momentum Cooper pairs which translates to the order parameter spatial variation. However, the dimer state can have in addition density modulation when interactions are present. } Moreover these dimer states are akin to other condensed matter structures used in the study of strongly interacting quantum liquids~\cite{Balents} and could be used as building blocks for simulating them.  As $|g_{\mathrm{eff}}|$ increases the steady state of the system will manifest this modulation in phase while having a homogeneous density distribution even in the absence of on-site interactions, all sites have the same mean atom number in Fig.\ref{pattdim}. The difference in phase that determines the  formation of the SFD state in the absence of interactions is shown in Fig.\ref{NID}.   

\begin{figure}[htbp!]
\begin{center}
\includegraphics[width=0.43\textwidth]{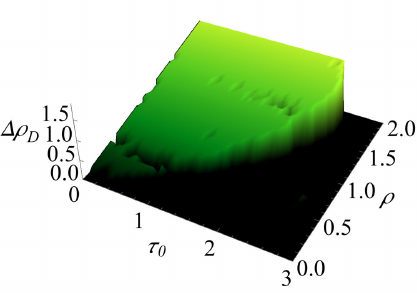}
\captionsetup{width=0.43\textwidth,justification=centerlast,font=small}
\caption{ 
[color on-line] Population difference between dimers $\Delta\rho_D=|\rho_A-\rho_B|/4$  for maximal light scattering through the bonds ($J_B\neq0$, $g_{\mathrm{eff}}<0$) as a function of the average density per site $\rho$ and the effective tunneling amplitude $\tau_0=2zt_0/U$. Whenever $\Delta\rho_D\neq 0$ the system is in the supersolid dimer phase (SSD), otherwise the system is in the normal SF state. Dimer populations are defined as: $\rho_A=\rho_1+\rho_2$, $\rho_B=\rho_3+\rho_4$. Dimers have different phase $\Delta\phi\neq0$ in the SSD, $U\neq0$.  Parameters are: $J_B=0.05$, $g_{\mathrm{eff}}N_s=-25U$, $N_s=100$, $z=6$ (3D). {  In all simulations the structured ground state has several $E_R$ lower than the homogenous ground state, substructure arises due to numerics.}
}
\label{SSD}
\end{center}
\end{figure}
 
However, the inclusion of on-site interactions generate additional competition via density pattern modulations which translate in the emergence of additional DW order supported by the system. As on-site interactions  partially suppress quantum fluctuations on each mode,  this competes with the fact that the system wants to maximize light scattering by imprinting the phase pattern. Therefore,  it becomes energetically favorable to have density imbalance. This can be traced back to the delicate balance between the phase modulation and the coupling of different modes in the effective Hamiltonian. Thus, the system can condensate into a super-solid dimer (SSD) phase, see Fig\ref{NID}. In this case, the quantum many-body state is a superfluid with spatial phase pattern and density modulation. The dimers favor a density imbalanced state between them while having in addition a change in phase. Graphically this corresponds to the situation where  the population in each site of the dimer is the same, but populations between dimers are different, see Fig. \ref{pattdim}. As the strength of the effective light matter coupling increases $|g_{\mathrm{eff}}|$ for large on-site interactions the SSD appears, see Fig.\ref{SSD}. As the $U$ decreases from large values, there is a transition to the more familiar regular SF state. In contrast to density coupling, bond coupling suppresses insulating phases even for large $U$ because tunneling is effectively enhanced in a non-trivial way effectively increasing quantum fluctuations. In addition, we note that the ground state of the system is massively degenerate as the number of components of the unit cell has $4d$ equivalent configurations. The bond order can be measured directly via the photon signal, $\langle\hat a^\dagger\hat a\rangle\propto|\langle\hat B\rangle|^2$.  The density modulations can be accessed by probing with and initial imbalanced state, while looking at the time of flight signal as additional peaks will appear consistent with the wave-vectors of the crystalline order formed, or via extraction of the structure factor, quantum non-demolition schemes can also be used \cite{QNDSanpera}.

\begin{figure}[htbp!]
\begin{center}
\includegraphics[width=0.43\textwidth]{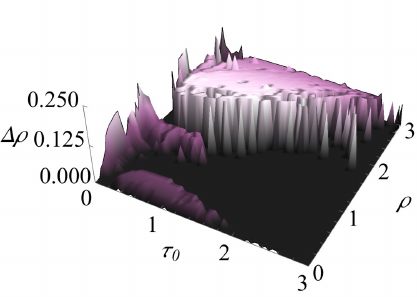}
\captionsetup{width=0.43\textwidth,justification=centerlast,font=small}
\caption{ 
[color on-line] Density wave order parameter $\Delta\rho$ for minimal light scattering through the bonds ($J_B\neq0$, $g_{\mathrm{eff}}>0$) as a function of the average density per site $\rho$ and the effective tunneling amplitude $\tau_0=zt_0/U$. Whenever $\Delta\rho\neq 0$ the system is in the supersolid phase (SS), otherwise the system is in the normal SF state.  Parameters are: $J_B=0.05$, $g_{\mathrm{eff}}N_s=25U$, $N_s=100$, $z=6$ (3D). { The substructure in the DW order parameter is due to numerics, in all simulations the structured ground state has energy that is several $E_R$ lower than the homogenous ground state.}
}
\label{NSS}
\end{center}
\end{figure}

As $g_{\mathrm{eff}}>0$, then light scatters minimally and instead of the formation of dimers, the phase modulation is absent. The quantum terms induced by light drive the behavior of the system. Then, the Hamiltonian collapses to two modes as the distinction between 4 modes is not necessary, as $\tilde{\eta}_{B,\varphi_\xi}=0$. This occurs because the energy is optimized without having a pattern between coherences. Thus, there will be no phase difference between coherences of nearest neighbors and dimer phases are strongly suppressed. We have instead for the simplified two mode Hamiltonian:
\begin{eqnarray}
\HH^b_{\mathrm{eff}}&=&\sum_{\xi=1}^2\HH^\xi_{\mathrm{eff}}
\\
\HH^\xi_{\mathrm{eff}}&\approx&\frac{N_s}{2}\big[z t_0\hat\beta_\xi-\mu\hat n_\xi+g_\mathrm{eff}|J_B|^2\delta S_{\varphi_\xi}^2
\nonumber\\
&+&
U \hat n_\xi(\hat n_\xi-1) 
\big]
\\
\hat\beta_\xi&=&
\big[
\psi_{\xi}^*\hat b^{\phantom{\dagger}}_{\xi+1}
+\psi_{\xi+1}^*\hat b_{\xi}^{\phantom{\dagger}}
+\psi_{\xi+1}^{\phantom{*}}\hat b_{\xi}^\dagger
+\psi_{\xi}^{\phantom{*}}\hat b_{\xi+1}^\dagger
\nonumber
\\
&-&
(\psi_{\xi}^*\psi^{\phantom{*}}_{\xi+1}
+\mathrm{c.c.})
\big]
\\
\delta S_{\varphi_\xi}^2&=&z\big[\hat b_\xi^{2\dagger}\hat b_{\xi+1}^{2\phantom{\dagger}}+\hat b_{\xi+1}^{2\dagger}\hat b_{\xi}^{2\phantom{\dagger}}
+2\hat n_{\xi}\hat n_{\xi+1}+\hat n_{\xi}+\hat n_{\xi+1}
\nonumber\\
&-&2(\psi^*_\xi\psi^{\phantom{*}}_{\xi+1}+\mathrm{c.c.})\hat \beta_\xi+(\psi^*_{\xi}\psi^{\phantom{*}}_{\xi+1}+\mathrm{c.c.})^2
\big],
\end{eqnarray}
where the component $\xi+2$ is the same as $\xi$. Here, the relevant contributions from the light matter coupling are the quantum fluctuations of the on-site coherences, the QOL. All the effect of the light-matter interaction reduces to the modification of quantum fluctuations of the $\hat B$ operator, which translate to the fluctuations in the order parameters. As the fluctuations in the oder parameter contain the term $\propto\hat n_1\hat n_2$, then this produces a DW instability. Depending on the value of light matter strength $g_{\mathrm{eff}}$ compared to the other parameters of the system, we can have a SS phase, see Fig \ref{NSS}. However, this SS phase is different from the density coupling case, as it only depends on the density pattern between nearest neighbors. This is a closer analogy to the conventional scenario of supersolidity~\cite{RevSS, DasSarma}. Different from maximal light scattering in either bonds or densities here, there is no coupling between all the coherences or sites of the lattice. The change in the Bose-Hubbard Hamiltonian is the remnant coupling induced by light, where the quantum fluctuations are maximized between short distances (up to nearest neighbors). In contrast to the maximal scattering case ($g_{\mathrm{eff}}<0$) the effective coupling strength $g_\mathrm{eff}$ needs to be larger by a factor of the number of illuminated sites to access a SS phase with large density imbalance. This is needed, to compensate for the fact that the effect of quantum fluctuations are just order $N_s$.  Still, the energy difference between the structured ground state and the competing homogenous ground state (i.e. regular SF) can be of several $E_R$.

\section{Light scattering different from $90^\circ$: Multi-component density orders}
As we can design the pattern of illumination via the projection of light onto the matter, it is possible to induce the formation of non-trivial density orders. By illuminating at an angle different from $90^\circ$ one changes the pattern of the coupling in the light induced effective interaction. As a particular example of what can be achieved and the relevant phases that emerge due to this, one can choose to illuminate with a traveling wave, such that the light projections between the cavity and the light pumped into the system are: $(\b{k}_{p}-\b{k}_c)\cdot\hat e_{\b{r}_j}=2\pi j/3$. This will induce $R=3$ spatial modes into the system, assuming that the lattice is sufficiently deep such that $\hat B\approx0$. We have then, that the light only couples to the density and the coupling is such that with mode operators $\hat D$ can written as:
\begin{equation}
\hat D=J_D(\hat N_{1}+e^{\frac{i2\pi}{3}}\hat N_{2}+e^{\frac{i4\pi}{3}}\hat N_{3}),
\end{equation}
where each mode corresponds to a third of the lattice sites ($N_s/3$). The effective Hamiltonian of the system can be written as:
\begin{eqnarray}
\HH^b_{\mathrm{eff}}&=&\sum_{\xi=1}^3\HH^\xi_{\mathrm{eff}}
\\
\HH^\xi_{\mathrm{eff}}&\approx&\frac{N_s}{3}\big[z t_0\hat\beta_\xi-\mu_\xi\hat n_\xi
+
U_\xi \hat n_\xi(\hat n_\xi-1)
\big]
\label{3mH}\\
\hat\beta_\xi&=&
\left[
(\psi_{\xi+1}^*+\psi_{\xi-1}^*)\hat b^{\phantom{\dagger}}_{\xi}
+(\psi_{\xi+1}^{\phantom{*}}+\psi_{\xi-1}^{\phantom{*}})\hat b_{\xi}^\dagger\phantom{\frac{1}{2}}\right.
\nonumber
\\
&-&\left.\frac{1}{2}
(\psi_{\xi}^*(\psi^{\phantom{*}}_{\xi+1}+\psi^{\phantom{*}}_{\xi-1})+\mathrm{c.c.})
\right]
\\
\mu_\xi&=&\mu-\frac{g_{\mathrm{eff}}N_s|J_D|^2}{3}\left(\rho_\xi-\frac{\rho_{\xi+1}+\rho_{\xi-1}}{2}\right)
\label{3mmu}
\\
U_\xi&=&U+2g_{\mathrm{eff}}|J_D|^2,
\end{eqnarray}
where the component $\xi+3$ is the same as $\xi$, similarly $\xi=0$ corresponds to $\xi=3$. From the above we can see that for minimal light scattering ($g_{\mathrm{eff}}>0$), the problem reduces to the Hamiltonian of the homogeneous system as density imbalance configurations are strongly suppressed and quantum fluctuations will shift the SF-MI transition as discussed previously. In the general case for $R$ modes we can expect that each of light induced modes decouple and their fluctuations depending on the light matter coupling can be affected independently.
\begin{figure}[t]
\begin{center}
\includegraphics[width=0.47\textwidth]{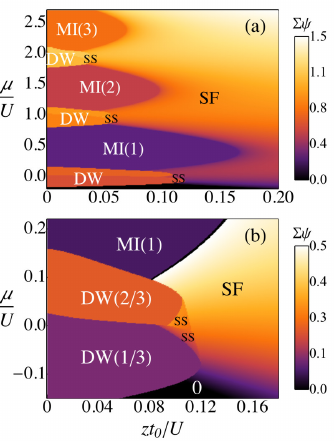}
\captionsetup{width=0.47\textwidth,justification=centerlast,font=small}
\caption{ 
[color on-line] (a) Phase diagram of multicomponent density ordered states, when geometry of the light pumped into the system such that 3 spatial modes can occur  ($R=3$). The system is SF, without any spatial pattern as the interaction is decreased (color on-line). The SF phase has total order parameter $\Sigma\psi=(|\psi_1|+|\psi_2|+|\psi_3|)/3\neq 0$ and $\psi_1=\psi_2=\psi_3=\psi$.  MI($n$) lobes appear for commensurate densities $n$, $\Sigma\psi=0$, $\rho=\rho_1=\rho_2=\rho_3$. In between MI lobes DW insulators form with $\langle\hat D\rangle\neq0$, maximal light scattering occurs then ($\Sigma\psi=0$). SS phases ($\Sigma\psi\neq 0$, $|\langle\hat D\rangle|^2\neq0$ ) are indicated near the boundary between DW and SF states. (b) Phase diagram closer to the DW-SS-SF transition at the tip of the DW insulators in between the first MI lobe. The particular composition of the DW insulators is revealed, as depending on the chemical potential one has different values of DW phases. The transition can have an intermediate SS state toward the homogeneous SF phase. Parameters are for (a) and (b): $J_D=1.0$, $J_B=0.0$, $g_{\mathrm{eff}}N_s/U=-0.5$, $N_s=100$, $z=6$ (3D). The grey scale [color] bar denotes $\Sigma\psi$ in the SF region, other different quantum phases (labeled) are denoted by different shades of grey [colors].}
\label{PD3m}
\end{center}
\end{figure}

 However, we see that when there is maximal light scattering ($g_{\mathrm{eff}}<0$) a DW instability occurs.  The coupling between adjacent mode favours density imbalance, effectively one has for the ground state energy terms  of the form $\propto\rho_{\xi}\rho_{\xi-1}>0$ and $\propto\rho_{\xi}\rho_{\xi+1}>0$. These arise from the effective chemical potential $\mu_\xi$ in Eq. (\ref{3mH}), that in fact depends on the density of the 3 modes in the system, Eq. (\ref{3mmu}). In contrast to illuminating at $90^\circ$, here a self-organized state with 3 components occurs. The state is 6 fold degenerate (a multi-component ``Schr\" odinger cat state"), thus the light amplitude gives the information about the DW order formation. The light amplitude is proportional to $\langle\hat D\rangle$, we have that when, $|\langle\hat D\rangle|^2=N_s^2|J_D|^2(\rho_1^2+\rho_2^2+\rho_3^2-\rho_1\rho_2-\rho_2\rho_3-\rho_1\rho_3)/9>0$ then the system maximizes light scattering and DW order is stablished. Depending on the competition between the light induced interaction and the atomic on-site interaction we will have that the system will support DW insulators, MI insulators and even SS states, see Fig.\ref{PD3m}. The DW occur as one of the three components is strongly suppressed while the remaining two are uneven, for strong interaction and fulling of non integer multiples  of 1/3 filling, e.g. $1/3,2/3,4/3,5/3,$ etc. These DW insulators appear in between MI lobes as the chemical potential is increased and have a critical value of $zt_0/U$ below the SF-MI transition.  As the system moves from the DW towards the SF state for small $U$ at fixed chemical potential, the system transitions via formation of  SS states, Fig.\ref{PD3m}.  The SS states we find, occur at the tip of the DW insulators, Fig.\ref{PD3m} (b).  It is interesting to note that the half-filled case will aways be in the SS state for large interaction as this state will be in the tip of the DW lobe, and it will shift to have a large (2/3) or small (1/3) DW while increasing $|g_{\mathrm{eff}}|$. There will be an intermediate region where the state will be better described as a mixed state of both configurations, all this consistent with the $t_0/U=0$ limit and exact diagonalization simulations~\cite{PRL2015}. The density pattern that emerges in the system has a period of 3 in units of the lattice spacing. In the case of other angles such that $R>3$ the behaviour would be similar, but now instead  DW insulating states and SS states would have period $R$ and the particular hierarchy of components and details in the competition between phases will be more complex as already foreseen in~\cite{PRL2015}.

\section{Conclusions}

Quantum light induces in its steady state an effective structured long range interaction. This steady-state effective many-body interaction changes the energetic landscape at the quantum level because atoms see a different energy landscape that depends on the cavity-pump detunings and the chosen spatial arrangement of the cavities and light pumped into the system. The atoms see a different energy landscape because the local energy at the sites is different in the illuminated regions. Thus atoms will tend to occupy or avoid those regions in space depending on the detunings. The cavity pump detunings determine the mechanism that will drive the symmetry breaking.  Moreover, depending on the choice of the pattern of the light with respect to the classical optical lattice the tunneling between the sites can be maximized or suppressed depending on the Wannier functions spatial overlap with the modes enhanced in the cavity. The interplay between the spatial distribution of the atoms given by the classical optical lattice and its characteristic Wannier functions with the spatial structure of the light modes pumped into the cavity determines the structure of the site energies. Thus scattering is maximized or minimized,  changing the local energies enhancing  different atomic processes. In turn the change in the light scattering from the atoms modifies the local energies and vice-versa, leading to a self-healing mechanism.  The change in the local energies  which couple to the density and coherence derived terms that couple to the tunneling amplitudes compete.  Depending on the light structure and this competition  the overall energy landscape seen by the matter is set. When light is scattered minimally by the atoms the formation of a quantum optical lattice occurs and quantum fluctuations induced by light determine the modification or emergence of new phases in the system. If on the contrary, light scatters maximally a dynamical optical lattice will form where the self-organization mechanism will be precursor of new phases.  This gives the opportunity via the structure of the light to modify by design the quantum many-body steady state of the matter which now has additional dependency via the quantum light. We have shown that the competition of additional processes due to the quantum light induced interaction with the regular processes (tunneling) and on-site repulsion in the Bose-Hubbard Hamiltonian opens the possibility for the atoms to condense to new phases that are energetically favorable. Among these new phases, we have found: states with limited quantum fluctuations, density waves, supersolids, bosonic dimer states, multi-component density waves and multicomponent supersolids.  Thus, the system can support all these additional spatially structured many-body quantum phases besides from the superfluid and Mott insulator homogeneous phases. The competition between global and local processes is the reason of the change in the phase diagram of the system. Therefore, previously energetically unfavorable configurations, thus with a small probability of occurring, become stable possible configurations. This however, constraints the energetics to be on a certain region of the Hilbert space. In turn,  the quantum nature of the atoms can give massively degenerate sets of states that are equally likely, depending on how many symmetries in the system are broken. As symmetries are broken, the degeneracy gets suppressed and the number of equivalent configurations becomes smaller. Thus, the emergence of structured quantum many-body phases occurs either via self-organization for DOL or modification of quantum fluctuations for QOL. 

Moreover, we have shown that the structure imprinted by light onto the matter by design can be exploited to simulate long-range interaction by light induced mode interactions. Certainly, this opens an additional venue of exploration towards quantum simulations. The effective mode Hamiltonians can be an acceptable representation of an otherwise experimentally hard to achieve quantum degenerate system with finite range interaction.  Our developments can be applied to other systems where light-matter coupling can be enhanced and the structure of it designed by cavities and external probes, as the scheme relies on off-resonant scattering. These results can be used in multi-mode optomechanical systems~\cite{RMP2014Optomech}, and other arrays of naturally occurring or synthetic quantum degenerate systems such as, spins, fermions, molecules (including biological ones)~\cite{LPhys2013}, ions~\cite{Ions2012} , atoms in multiple cavities~\cite{ArrayPolaritons2006}, semiconductor~\cite{SQubits2007}  or superconducting qubits~\cite{JCQubits2009}.  
   
\section*{Acknowledgements}
This work was supported by the EPSRC (Grant No. EP/I004394/1).

  \end{document}